%% file: neurips_2026.tex
\documentclass{article}

\PassOptionsToPackage{numbers,compress}{natbib}

 \usepackage[preprint]{neurips_2026}

\usepackage[utf8]{inputenc} 
\usepackage[T1]{fontenc}    
\usepackage{hyperref}       
\usepackage{url}            
\usepackage{booktabs}       
\usepackage{amsfonts}       
\usepackage{nicefrac}       
\usepackage{microtype}      
\usepackage[table]{xcolor}   
\usepackage{microtype}
\usepackage{enumitem}
\usepackage{bbm}
\usepackage{graphicx}
\usepackage{pifont} 
\usepackage{booktabs} 
\usepackage{ragged2e}
\usepackage{booktabs,makecell, multirow,tabularx}
\usepackage{algorithm}
\usepackage{algorithmic}

\usepackage{amsmath}
\usepackage{amssymb}
\usepackage{mathtools}
\usepackage{amsthm}
\theoremstyle{plain}
\newtheorem{theorem}{Theorem}[section]

\theoremstyle{definition}
\newtheorem{definition}[theorem]{Definition}

\theoremstyle{remark}
\newtheorem{remark}[theorem]{Remark}
\usepackage{caption}
\usepackage{subcaption} 
\usepackage{cleveref} 
\usepackage{mdframed}
\usepackage{wrapfig}
\newenvironment{promptbox}{%
  \begin{mdframed}[
    linewidth=0.4pt,
    linecolor=black!12,
    backgroundcolor=gray!10,
    innerleftmargin=8pt,
    innerrightmargin=8pt,
    innertopmargin=6pt,
    innerbottommargin=6pt,
    skipabove=\smallskipamount,
    skipbelow=\smallskipamount
  ]
  \small
}{\end{mdframed}}

\usepackage{xspace}
\newcommand{\AttackName}{\texttt{GDA}\xspace}

\AtBeginEnvironment{equation}{\small}
\AtBeginEnvironment{equation*}{\small}
\AtBeginEnvironment{align}{\small}
\AtBeginEnvironment{align*}{\small}
\AtBeginEnvironment{gather}{\small}
\AtBeginEnvironment{gather*}{\small}
\AtBeginEnvironment{multline}{\small}
\AtBeginEnvironment{multline*}{\small}

\title{Grounding-Driven Attack: Improving Encoder-based Adversarial Transferability against Large Vision-Language Models}

\author{%
\begin{tabular}{c}
Xinwei Zhang$^{1}$, Li Bai$^{1}$, Tianwei Zhang$^{2}$, Youqian Zhang$^{1}$, Qingqing Ye$^{1}$\\
Yingnan Zhao$^{3}$, Ruochen Du$^{3}$, Haibo Hu$^{1 *}$\\
{\small $^{1}$The Hong Kong Polytechnic University, Hong Kong, China}\\
{\small $^{2}$Nanyang Technological University, Singapore}\\
{\small $^{3}$Harbin Engineering University, Harbin, China}\\
{\small \texttt{xin-wei.zhang@connect.polyu.hk, haibo.hu@polyu.edu.hk}}\\
{\small $^{*}$Corresponding author.}
\end{tabular}
}

\begin{document}

\maketitle

\begin{abstract}
Large vision-language models (LVLMs) have achieved impressive performance across multimodal tasks, but their reliance on visual inputs exposes them to adversarial threats. 
Encoder-based attacks provide an efficient alternative to end-to-end optimization by crafting perturbations through the vision encoder alone. 
However, existing encoder-based attacks often assume that the surrogate encoder is identical or similar to the victim LVLM's vision encoder. 
In this work, we present a systematic study of their transferability in more realistic black-box deployments with heterogeneous LVLM architectures.
We find that model-specific visual evidence is inconsistent across models, whereas text-conditioned grounding regions are more closely tied to caption-relevant evidence and provide a more stable transfer target. 
However, existing attacks remain weakly aligned with and insufficiently disrupt these regions.
Motivated by these findings, we propose Grounding-Driven Attack (\AttackName), which aligns perturbation optimization with text-grounded evidence. 
\AttackName combines Grounding-Aware Perturbation Allocation to concentrate perturbation budget on grounded evidence regions with Grounding-Centric Evidence Disruption to intensify their global and local disruption. 
Experiments across diverse victim models and tasks show that \AttackName consistently outperforms existing encoder-based attacks in black-box transfer. 
These results highlight the central role of text-grounded evidence in adversarial transferability and motivate grounding-aware robustness evaluation and defense design.
\end{abstract}

\setlength{\textfloatsep}{8pt plus 1pt minus 2pt}
\setlength{\floatsep}{6pt plus 1pt minus 2pt}
\setlength{\intextsep}{6pt plus 1pt minus 2pt}
\setlength{\abovecaptionskip}{4pt plus 1pt minus 1pt}
\setlength{\belowcaptionskip}{0pt}
\setlength{\abovedisplayskip}{7pt plus 1pt minus 2pt}
\setlength{\belowdisplayskip}{7pt plus 1pt minus 2pt}
\setlength{\abovedisplayshortskip}{5pt plus 1pt minus 2pt}
\setlength{\belowdisplayshortskip}{5pt plus 1pt minus 2pt}
\captionsetup{skip=4pt}

\input{Section/1_Intro}

\input{Section/2_Preliminaries}
\input{Section/4_challenges}

\input{Section/5_Method}

\input{Section/6_Experiment}

\section{Conclusion}
\label{sec:conclusion}
In this paper, we study encoder-based adversarial transferability against LVLMs in zero-query black-box settings. 
Our analysis reveals a key bottleneck of existing attacks: model-specific visual evidence is unstable across heterogeneous LVLMs, while text-grounded evidence is more transferable but remains insufficiently disrupted. 
Motivated by this insight, we propose \AttackName, which aligns perturbation optimization with text-grounded visual evidence through grounding-aware perturbation allocation and global-local disruption. 
Experiments across diverse LVLMs, tasks, surrogate models, and commercial systems show that \AttackName consistently improves black-box transferability while preserving good perceptual quality. Our results highlight text-grounded visual evidence as a useful perspective for understanding and evaluating LVLM security in realistic heterogeneous settings.
By revealing the role of grounding in adversarial transfer, our study provides a useful foundation for developing both stronger attacks and future defenses that account for cross-model grounding evidence.

\bibliographystyle{unsrtnat}
\bibliography{Myref}


\appendix

\setlength{\textfloatsep}{12pt plus 2pt minus 4pt}
\setlength{\floatsep}{12pt plus 2pt minus 4pt}
\setlength{\intextsep}{12pt plus 2pt minus 4pt}
\setlength{\abovecaptionskip}{6pt plus 1pt minus 1pt}
\setlength{\belowcaptionskip}{0pt}
\setlength{\abovedisplayskip}{10pt plus 2pt minus 4pt}
\setlength{\belowdisplayskip}{10pt plus 2pt minus 4pt}
\setlength{\abovedisplayshortskip}{6pt plus 2pt minus 3pt}
\setlength{\belowdisplayshortskip}{6pt plus 2pt minus 3pt}
\captionsetup{skip=6pt}

\input{Section/Appendix}



\end{document}

%% file: Section/1_Intro.tex
\section{Introduction}
With the rapid advancement of data scale, computational resources, and model architectures, large language models (LLMs) have demonstrated impressive capabilities in understanding and generating natural language. Building upon the semantic reasoning capabilities of LLMs, large vision-language models (LVLMs), such as GPT-4V \cite{openai2024gpt4technicalreport} and Gemini \cite{team2023gemini}, incorporate visual inputs to support multiple modalities, thereby significantly enhancing instruction following and user interaction in complex vision-language scenarios.
However, the visual modality is inherently more susceptible to imperceptible perturbations compared to the textual modality~\cite{qiu2022benchmarking,zhaoEvaluatingAdversarialRobustness2023}, which could amplify the vulnerability of the LVLM to adversarial examples. This raises serious security concerns for deploying LVLMs in safety-critical applications, such as medical image analysis \cite{nath2025vila,lin2025healthgpt} and autonomous systems \cite{lubberstedt2025v3lma}.

Researchers have proposed a variety of attacks against LVLMs by generating adversarial vision input. A straightforward attack strategy is end-to-end optimization \cite{schlarmannAdversarialRobustnessMultiModal2023}, where perturbations are directly crafted based on the model’s final output. This approach is typically model-specific and computationally expensive since it requires access to the full forward and backward pass of the entire LVLM~\cite{cuiRobustnessLargeMultimodal2024}.
As a more lightweight alternative, recent efforts have shifted toward encoder-based attacks, which perturb the input image by targeting only the vision encoder of the LVLM \cite{cuiRobustnessLargeMultimodal2024,wangBreakVisualPerception2024,dongHowRobustGoogles2023,tuHow2024}. 
These methods leverage the property of \textit{encoder-based transferability}, i.e., adversarial perturbations effective on a surrogate vision encoder are expected to remain effective in different victim LVLMs.
In most studies, this property relies on the strong assumption \cite{wangBreakVisualPerception2024,cuiRobustnessLargeMultimodal2024,wangInstructTAInstructionTunedTargeted2024, xieChainAttackRobustness2025} that the victim LVLM shares the same or a highly similar vision encoder with the surrogate. 
However, the effectiveness of adversarial attacks against LVLMs with heterogeneous architectures in encoders and language modules, which reflects a more realistic scenario \cite{chen2024internvl1,openai2024gpt45,kimiteam2025kimivltechnicalreport}, is less explored.

In this work, we present a systematic study of encoder-based adversarial transferability in LVLMs.
We investigate the poor cross-model transfer of existing encoder-based attacks through grounding metrics, masking studies, and attack heatmaps.
Our analysis reveals that model-specific visual evidence is inconsistent across heterogeneous LVLMs: regions highlighted by the surrogate encoder often differ from those used by the victim, reducing the transferability of attacks based on surrogate gradients or attention signals.
We further show that text-conditioned grounding provides a more reliable target: these regions are more closely tied to caption-relevant evidence, and at the same masking budget, removing them degrades caption content much more than random removal.
However, existing encoder-based attacks are only weakly aligned with these regions and often fail to disrupt them sufficiently, as perturbation mass either drifts to transfer-weak background areas or remains too sparse over evidence-carrying patches.


Motivated by these findings, we propose \textbf{Grounding-Driven Attack (\AttackName)}, a framework designed to enhance adversarial transferability across LVLMs. Since text-grounded evidence regions should be prioritized and existing attacks still fail to localize and disrupt them effectively, \AttackName comprises two complementary components:
\textit{(i) Grounding-Aware Perturbation Allocation} improves perturbation localization by anchoring updates to text-conditioned, evidence-rich patches derived from the image description and down-weighting unstable background regions that contribute little transferable signal.
\textit{(ii) Grounding-Centric Evidence Disruption} strengthens perturbation effectiveness on those grounded regions at both the global level in the joint embedding space and the local level over noun-phrase-aligned patches. Across multiple vision--language tasks and a heterogeneous black-box setting spanning both open-source and commercial LVLMs, \AttackName consistently improves transferability over existing encoder-based attacks while preserving comparable perceptual quality.


In summary, our contributions are:
\ding{182} We investigate encoder-based adversarial transferability in LVLMs under a realistic heterogeneous black-box setting.
\ding{183} We identify a grounding-level source of poor transferability: model-specific visual evidence is inconsistent across LVLMs, while text-conditioned grounding regions are more stable, caption-relevant, and insufficiently disrupted by existing attacks.
\ding{184} We introduce \AttackName, a grounding-driven attack framework that combines grounding-aware perturbation allocation and grounding-centric evidence disruption to localize and intensify perturbations on text-grounded evidence regions.
\ding{185} We demonstrate the effectiveness of \AttackName across multiple tasks and diverse open-source and commercial LVLMs, showing stronger black-box transferability than existing encoder-based attacks.

%% file: Section/2_Preliminaries.tex
\section{Preliminary}
\label{sec:preliminaries}
\noindent\textbf{Large Vision-Language Models.}
A typical LVLM $F_\theta$ consists of three components: 
a vision encoder $f_\phi$ that extracts patch-level visual features, a modality projector $M_\psi$ that maps visual embeddings into the textual embedding space, and a large language model $L_\tau$ that performs text generation. 
Formally, given an image $I$ and a text prompt $T$, the LVLM output $y$ is computed as
$y = F_\theta(I, T) = L_\tau(M_\psi(f_\phi(I)), T)$.
LVLMs mainly differ in the design of these three components, and this architectural heterogeneity also complicates adversarial transferability across LVLMs.

\noindent\textbf{Adversarial Examples against LVLMs.}
Adversarial examples add imperceptible perturbations to inputs to induce incorrect model behaviors~\cite{demontis2019adversarial,liBlack2022}. 
For LVLMs, perturbations on the image can propagate through the vision--language pipeline and alter the generated text~\cite{cuiRobustnessLargeMultimodal2024,wangBreakVisualPerception2024}. 
Since images are easier to manipulate in practice and are highly susceptible to subtle perturbations, most LVLM attacks perturb the image while keeping the prompt unchanged~\cite{cuiRobustnessLargeMultimodal2024,wangBreakVisualPerception2024,dongHowRobustGoogles2023,tuHow2024,zhaoEvaluatingAdversarialRobustness2023}. 
Formally, given a clean image $I$ and a fixed prompt $T$, the goal is to find $\delta$ such that
\begin{equation}
F_\theta(I+\delta, T) \ne y,\quad \text{s.t. } \|\delta\|_p \le \epsilon .
\end{equation}

Existing attacks mainly differ in which part of the model they optimize against.
\textbf{End-to-end attacks} optimize $\delta$ to directly change the final LVLM outputs~\cite{dongHowRobustGoogles2023,schlarmannAdversarialRobustnessMultiModal2023}:
$\max_{\|\delta\|_p \leq \epsilon} 
\mathcal{L}\!\left(F_\theta(I+\delta, T),\, F_\theta(I, T)\right)$, where $\mathcal{L}$ measures the output discrepancy (e.g., cross-entropy or contrastive loss).
\textbf{Encoder-based attacks} instead optimize $\delta$ by disrupting intermediate visual representations extracted by a surrogate encoder $f_\phi$~\cite{cuiRobustnessLargeMultimodal2024,wangBreakVisualPerception2024,zhaoEvaluatingAdversarialRobustness2023,dongHowRobustGoogles2023,tuHow2024}:
\begin{equation}
\max_{\|\delta\|_p \leq \epsilon} 
\mathcal{L}\!\left(f_\phi(I+\delta),\, f_\phi(I)\right),
\end{equation}
where $\mathcal{L}$ is typically a feature distance loss (e.g., cosine distance).
Encoder-based attacks have recently become attractive because they avoid optimizing over the full LVLM and do not require access to the generated outputs, making them computationally cheap~\cite{cuiRobustnessLargeMultimodal2024}. 

\noindent\textbf{Encoder-based Transferability.}
Building on the transferability concept studied in \cite{liu2017delving,demontis2019adversarial,xiaEnhance2024}, we define encoder-based transferability at the instance level as follows.
\begin{definition}[Encoder-based Transferability]
Let the victim LVLM be $V \coloneqq \{f_{\phi}^{V}, M_{\psi}^{V}, L_{\tau}^{V}\}$,
and $S \coloneqq f_{\phi}^{S}$ be a surrogate vision encoder.
Given a normal instance consisting of an image $I$, a prompt $T$, and a label $y$,
let $I + \delta$ be a perturbed image crafted against a surrogate vision encoder $S$.
The \emph{encoder-based transferability $\mathcal{T}$} from $S$ to $V$ at the instance $(I, T, y)$ is defined as
\[
\mathcal{T}(S \to V; I, T, y)
= \mathbb{I}\Big[
L_{\tau}^{V}\big(M_{\psi}^{V}(f_{\phi}^{V}(I+\delta)), T\big) \ne y
\Big],
\]
where $\mathbb{I}[\cdot]$ is the indicator function.
\label{definition}
\end{definition}
Based on the above definition, achieving encoder-based transferability in LVLMs requires satisfying two conditions.
\ding{182} \textbf{Cross-encoder transferability:} the perturbations $\delta$ crafted against a surrogate encoder $f_\phi^S$ must also alter the victim LVLM’s vision encoder $f_\phi^V$, even when their architectures differ (e.g., patch size or pretraining data).
\ding{183} \textbf{Encoder-to-model transferability:} once the perturbation $\delta$ has successfully transferred to the victim encoder $f_\phi^V$, the resulting representation must further propagate through the alignment and language generation modules of the victim LVLM to ultimately alter its output.
These two sub-conditions jointly determine the practical transferability of encoder-based attacks across the full LVLM inference.


\noindent\textbf{Threat Model.}
We consider a zero-query black-box threat model for studying encoder-based transferability. The adversary has no access to the victim LVLM's internal components, including its architecture, parameters, vision encoder, or output probabilities. Instead, it leverages a publicly available vision--language pre-trained model (e.g., CLIP), which comprises a vision encoder $f_v(\cdot)$ and a text encoder $f_t(\cdot)$ jointly trained to align image--text pairs in a shared $d$-dimensional embedding space. For clarity, we denote $f_v(\cdot)$ as the surrogate vision encoder $f_\phi^S(\cdot)$ in line with Definition~\ref{definition}. Adversarial examples are crafted solely on this surrogate, without querying the victim LVLM, and are then directly transferred to the victim model. Unlike query-based attacks that rely on repeated interactions with the target model to approximate gradients, this setting removes such dependencies and better reflects realistic transfer-based black-box attacks. We provide detailed attack scenarios with a case study in Appendix~\ref{sec:case_study}.

%% file: Section/4_challenges.tex
\section{Empirical Analysis of Transferability Limitations}
\label{sec:analysis}

\begin{figure*}[t]
    \centering

    \begin{minipage}[t]{0.39\textwidth}
        \vspace{0pt}
        \centering
        \captionsetup{type=table}
        \caption{Grounding consistency across models.}
        \label{tab:spatial_misalignment_six_row}

        \vspace{-1mm}
        \scriptsize
        \setlength{\tabcolsep}{4pt}
        \renewcommand{\arraystretch}{0.95}
        \resizebox{\linewidth}{!}{%
        \begin{tabular}{@{}lcc@{}}
            \toprule
            \textbf{Comparison} & \textbf{IoU@Top20\%} & \textbf{Spearman $\rho$} \\
            \midrule
            SigLIP / CLIP-L      & $0.156 \pm 0.040$ & $0.157 \pm 0.109$ \\
            SigLIP / CLIP-B      & $0.095 \pm 0.038$ & $-0.067 \pm 0.163$ \\
            SigLIP / Grad-ECLIP  & $\mathbf{0.164 \pm 0.068}$ & $\mathbf{0.206 \pm 0.161}$ \\
            \midrule
            LLaVA / CLIP-L       & $0.112 \pm 0.034$ & $0.032 \pm 0.132$ \\
            LLaVA / CLIP-B       & $0.109 \pm 0.054$ & $0.024 \pm 0.179$ \\
            LLaVA / Grad-ECLIP   & $\mathbf{0.223 \pm 0.090}$ & $\mathbf{0.325 \pm 0.180}$ \\
            \bottomrule
        \end{tabular}%
        }
    \end{minipage}
    \hfill
    \begin{minipage}[t]{0.57\textwidth}
        \vspace{0pt}
        \centering
        \captionsetup{type=figure}
        \begin{subfigure}[t]{0.32\linewidth}
            \centering
            \includegraphics[width=\linewidth]{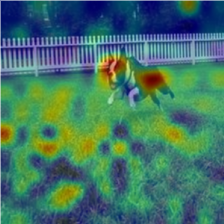}
            \caption{CLIP ViT-L/14.}
            \label{subfig:e1}
        \end{subfigure}
        \hfill
        \begin{subfigure}[t]{0.32\linewidth}
            \centering
            \includegraphics[width=\linewidth]{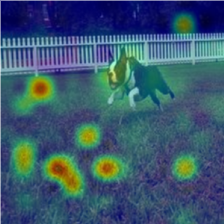}
            \caption{CLIP ViT-B/16.}
            \label{subfig:e2}
        \end{subfigure}
        \hfill
        \begin{subfigure}[t]{0.32\linewidth}
            \centering
            \includegraphics[width=\linewidth]{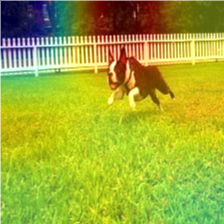}
            \caption{LLaVA-v1.5-7b.}
            \label{subfig:e3}
        \end{subfigure}
        \caption{Attention maps across models.}
        \label{fig:challenge_example1}
    \end{minipage}
\end{figure*}

\begin{figure*}[t]
    \centering

    \begin{minipage}[t]{0.39\textwidth}
        \vspace{0pt}
        \centering

        \captionsetup{type=table}
        \caption{VT-Attack vs. grounded regions.}
        \label{tab:grad_eclip_vt_only}

        \vspace{-1mm}
        \scriptsize
        \setlength{\tabcolsep}{5pt}
        \renewcommand{\arraystretch}{0.95}
        \resizebox{\linewidth}{!}{%
        \begin{tabular}{lc}
            \toprule
            \textbf{Metric} & \textbf{VT-attack} \\
            \midrule
            $\mathrm{Spearman}$ $\uparrow$
            & $0.0524 \pm 0.0834$ \\
            $\mathrm{IoU@Top20\%}$ $\uparrow$
            & $0.1303 \pm 0.0443$ \\
            \midrule
            $\mathrm{coverage@Top20\%\_sal}$ $\uparrow$
            & $0.4108 \pm 0.0868$ \\
            \bottomrule
        \end{tabular}%
        }
    \end{minipage}
    \hfill
    \begin{minipage}[t]{0.57\textwidth}
        \vspace{0pt}
        \centering

        \captionsetup{type=figure}
        \begin{subfigure}[t]{0.32\linewidth}
            \centering
            \includegraphics[width=\linewidth]{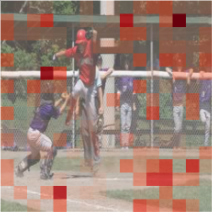}
            \caption{}
            \label{subfig:why_a}
        \end{subfigure}
        \hfill
        \begin{subfigure}[t]{0.32\linewidth}
            \centering
            \includegraphics[width=\linewidth]{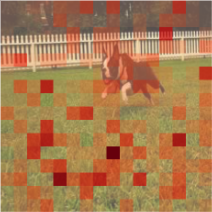}
            \caption{}
            \label{subfig:why_b}
        \end{subfigure}
        \hfill
        \begin{subfigure}[t]{0.32\linewidth}
            \centering
            \includegraphics[width=\linewidth]{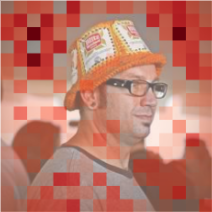}
            \caption{}
            \label{subfig:why_c}
        \end{subfigure}
        \caption{VT-Attack perturbation heatmaps.}
        \label{fig:clip_failure}
    \end{minipage}

    \vspace{-3mm}
\end{figure*}
In this section, we present an empirical study of the transferability limitations of existing encoder-based attacks across heterogeneous LVLMs. We summarize three empirical findings that characterize recurring failure patterns and motivate the design of our method.

\noindent\textbf{$\bullet$ Finding 1: Model-specific visual evidence is inconsistent across heterogeneous LVLMs, while text-conditioned grounding provides a better-aligned attack target.}

Figure~\ref{fig:challenge_example1} illustrates this phenomenon with attention maps from different encoders and one LLaVA model. We observe inconsistencies at two levels.
\textbf{(1) Across vision encoders.}
By comparing \Cref{subfig:e1} and \Cref{subfig:e2}, we find that different encoders assign high importance to different regions of the same image. One encoder distributes attention over multiple regions, while the other focuses on a smaller set of patches. This weakens \textit{cross-encoder transferability}, since perturbations optimized for one encoder may not affect the regions emphasized by another.
\textbf{(2) Between the encoder and the LVLM.}
By comparing \Cref{subfig:e1,subfig:e3}, we further observe that the regions highlighted by the encoder alone differ from those emphasized by LLaVA during generation. This suggests that encoder-sensitive patches are not necessarily the visual evidence used by the LLM to produce the final response, thereby weakening \textit{encoder-to-model transferability}.

Table~\ref{tab:spatial_misalignment_six_row} further provides a quantitative comparison. We measure consistency using IoU@Top20\%, which computes the overlap between the top 20\% most important patches of two maps, and Spearman's $\rho$, which measures the rank correlation between their patch-importance scores. Higher values indicate stronger agreement, while low IoU and weak rank correlation indicate that model-specific visual evidence is highly inconsistent across encoders and LVLMs.
Grad-ECLIP~\cite{zhaoGradientbasedVisualExplanation2024} produces a text-conditioned grounding map that localizes image regions most relevant to the image description. These Grad-ECLIP-derived regions show relatively better agreement with the LVLM, suggesting that text-conditioned grounding offers a more stable proxy for selecting transferable perturbation targets.

\begin{wrapfigure}[11]{r}{0.38\linewidth}
    \centering
    \vspace{-2mm}
    \includegraphics[width=\linewidth]{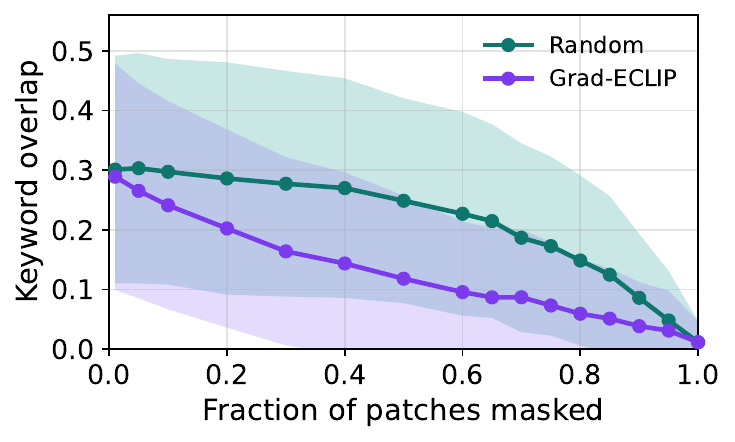}
    \caption{Random masking vs. grounded masking.}
    \label{fig:mask_guided_vs_random}
    \vspace{-2mm}
\end{wrapfigure}
\noindent\textbf{$\bullet$ Finding 2: Text-conditioned grounding regions contribute disproportionately to caption-relevant outputs.}

Beyond cross-model evidence inconsistency, not all visual patches contribute equally to LVLM generation.
Figure~\ref{fig:mask_guided_vs_random} compares random masking with Grad-ECLIP-guided masking. Here keyword overlap measures the fraction of salient content words shared between the generated caption before and after masking, so a larger drop indicates stronger grounding disruption.
At the same masking ratio, random masking causes only moderate degradation in keyword overlap, indicating that LVLMs can tolerate the removal of many arbitrary visual patches.
In contrast, masking patches selected by Grad-ECLIP leads to consistently larger drops.
This suggests that text-conditioned grounding regions contain disproportionately important evidence for preserving caption-relevant content, and therefore provide more promising perturbation targets than arbitrary or purely visually salient regions.

\medskip
\noindent\textbf{$\bullet$ Finding 3: Existing encoder-based attacks are weakly aligned with text-conditioned grounding regions and fail to disrupt them sufficiently.}

To understand how the above findings manifest in practice, we analyze VT-Attack~\cite{wangBreakVisualPerception2024} and visualize patch-level perturbation heatmaps in \Cref{fig:clip_failure}. Each red overlay denotes the cosine distance between clean and adversarial patch embeddings, so deeper red indicates larger feature deviations.
We observe two recurring failure patterns.
First, many perturbations are concentrated on visually salient but transfer-unstable background regions.
For example, in \Cref{subfig:why_c}, a large portion of the perturbation is allocated to the background rather than to the person's face.
Such regions are often inconsistent across encoders and are less likely to be used by the LVLM during generation, making the perturbations ineffective after transfer.
Second, even when perturbations partially overlap with object regions, they are often too weak or too sparse to substantially disrupt the most relevant caption-conditioned evidence.
As shown in \Cref{subfig:why_b}, the attack perturbs only a limited subset of patches on the dog, leaving large portions of its body intact.
Similarly, in \Cref{subfig:why_c}, the perturbation around the face is scattered and misses several critical features such as the eyes and mouth.
These examples suggest that merely touching relevant object regions is not sufficient; the disruption must also be concentrated over the evidence-carrying patches identified by text-conditioned grounding.

Table~\ref{tab:grad_eclip_vt_only} provides a quantitative view of this mismatch. Here Spearman and IoU@Top20\% again measure how well perturbation intensity aligns with grounding relevance, while coverage@Top20\%\_sal measures what fraction of the top 20\% most relevant Grad-ECLIP patches are actually touched by the perturbation.
VT-Attack shows weak alignment with text-conditioned grounding regions, with a low Spearman correlation of $0.0524$ and a low IoU@Top20\% of $0.1303$.
Its coverage of the top 20\% most relevant Grad-ECLIP regions is also limited to $0.4108 \pm 0.0868$.
These results confirm that existing encoder-based attacks do not sufficiently concentrate perturbations on the evidence-carrying patches used by the LVLM.

%% file: Section/5_Method.tex



\section{Grounding-Driven Attack}
\label{sec:SGMA}
\begin{figure}[t]
	\centering
\includegraphics[width=\columnwidth]{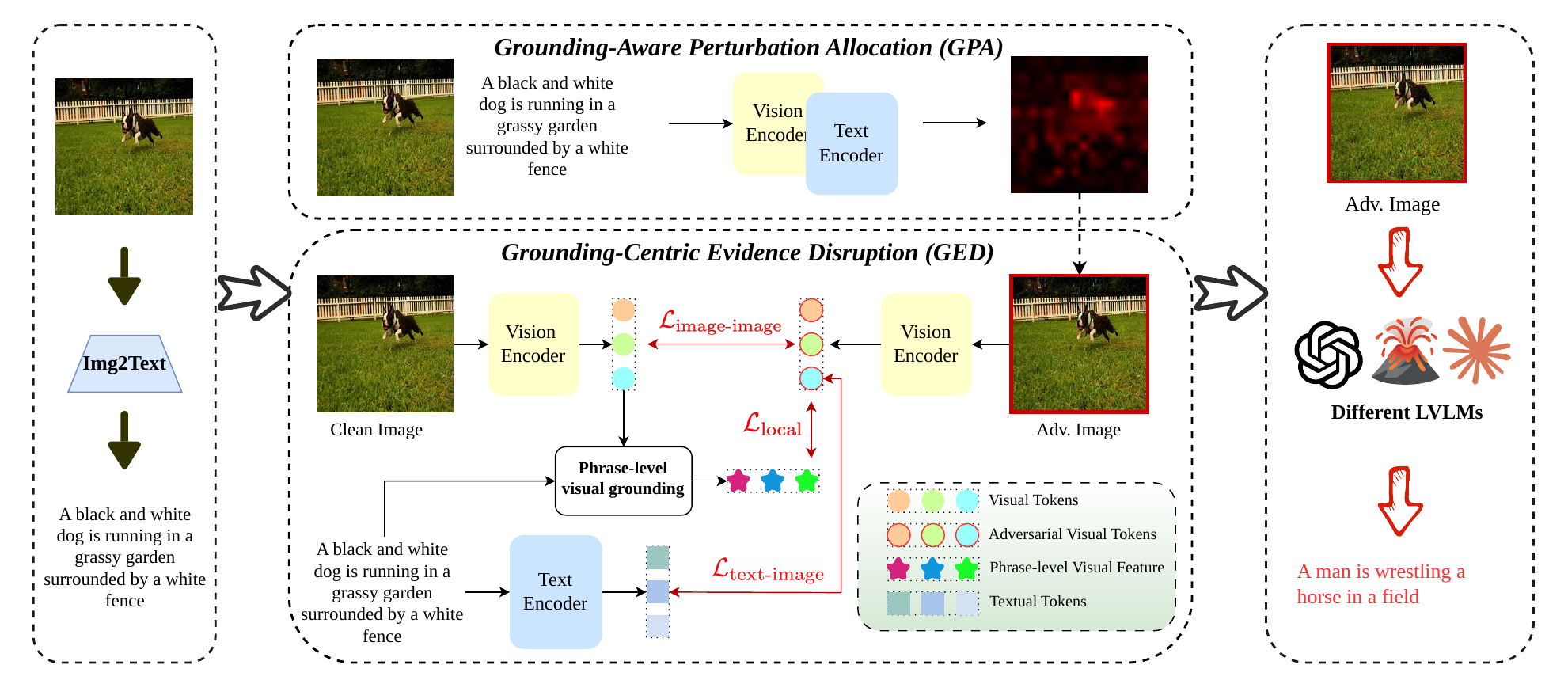}
	\caption{The framework of \AttackName.}
         \vspace{-3mm}
	\label{Fig:SGMA}
\end{figure}


Building on the above analysis, we propose \textbf{Grounding-Driven Attack (\AttackName)}. 
The core idea is to address two key questions for transferable perturbation optimization: \emph{where} the perturbation budget should be allocated and \emph{how strongly} the selected text-grounded regions should be disrupted. 
As shown in \Cref{Fig:SGMA}, \AttackName consists of two components. 
First, \textbf{Grounding-Aware Perturbation Allocation (GPA)} directs updates toward text-grounded, evidence-carrying regions while suppressing inconsistent background areas through a soft grounding mask derived from the paired description. 
Second, \textbf{Grounding-Centric Evidence Disruption (GED)} increases the strength and spatial density of perturbations within these grounded regions through both global and local disruption. 
We detail the two components below.

\subsection{Grounding-Aware Perturbation Allocation}
\label{sec:semantic_mask}
Following the above analysis, GPA allocates stronger perturbations to text-grounded foreground regions and suppresses unnecessary updates on inconsistent background areas. To achieve this, we adopt Grad-ECLIP~\cite{zhaoGradientbasedVisualExplanation2024}, which provides fine-grained localization of text-aligned visual tokens in CLIP. Unlike gradient- or attention-based techniques applied directly to encoders (e.g., Grad-CAM~\cite{selvaraju2017grad}, attention rollout~\cite{abnarQuantifyingAttentionFlow2020}), which often introduce unacceptable model-specific biases~\cite{wangFeatureImportanceawareTransferable2021,liImprovingTransferabilityAdversarial2024}, Grad-ECLIP leverages the text embedding as an anchor to constrain gradients, enabling cross-modal localization of text-aligned regions. Although these regions may not perfectly match the important grounding regions of other LVLMs, CLIP’s large-scale pretraining on diverse image–text pairs ensures they serve as a relatively stable proxy for transferable perturbation targets.
Concretely, GPA includes the following steps.

\noindent\textbf{Step 1: Grounding Reference Generation.}
We first query a pretrained image captioning model (e.g., GPT-4o) to summarize the image $I$ into a concise caption $T_d$. This summary $T_d$ captures the main visual content of the image and serves as a natural reference for grounding. By aligning $I$ with $T_d$, we can highlight the visual regions contributing most to that grounded content. We further verify in Appendix~\ref{sec:additional_ablaction_study} (Table~\ref{tab:caption_quality}) that replacing the captioner has only a negligible effect on attack performance, indicating that this step mainly requires a coarse description of the dominant visual entities rather than a highly optimized LVLM caption.

\noindent\textbf{Step 2: Text-Grounded Localization.}
We then leverage Grad-ECLIP to derive a text-conditioned patch-level attention map $\mathbf{A}$ for $I$.  
Given the normalized image and text embeddings $\mathbf{v} = f_v(I)$ and $\mathbf{t} = f_t(T_d)$, we compute their cosine similarity $\mathcal{L}_{\text{cos}} = \mathbf{v}^\top \mathbf{t}$ and backpropagate it to obtain the gradient $\nabla_{\mathbf{o}_{\text{CLS}}} \mathcal{L}_{\text{cos}}$ with respect to the [CLS] token output $\mathbf{o}_{\text{CLS}}$. 
Let $\mathbf{v}_i \in \mathbb{R}^{d_v}$ denote the value vector of the $i$-th image patch token in the final transformer layer (where $d_v$ is the vision encoder's hidden dimension), and $\alpha_i$ the averaged attention weight assigned to patch $i$ by the [CLS] token across all attention heads. The saliency score for patch $i$ is then calculated as $s_i = \alpha_i \cdot \left( \mathbf{v}_i \cdot \nabla_{\mathbf{o}_{\text{CLS}}} \mathcal{L}_{\text{cos}} \right)$,
which reflects the patch’s contribution to the similarity, modulated by both attention and gradient relevance.

The patch-level scores $\{s_i\}_{i=1}^{HW}$ (with $H \times W$ denoting the patch grid size) are then reshaped into a 2D spatial map $\mathbf{A} \in \mathbb{R}^{H \times W}$ and bilinearly upsampled to the input resolution $R \times R$, 
yielding a pixel-level grounding mask 
$\mathbb{M} \in [0,1]^{R \times R}$.
This map captures the patch-level gradient attribution of visual features with respect to the text prompt and serves as a grounding mask for perturbation localization.


\noindent\textbf{Step 3: Perturbation Allocation.}
Finally, we employ $\mathbb{M}$ to generate a pixel-wise perturbation map $\boldsymbol{\epsilon}_{\text{map}} \in \mathbb{R}^{R \times R}$ under a global perturbation budget constraint, where the total perturbation budget is bounded while allowing adaptive allocation across pixels. 
We define $\epsilon > 0$ as the average per-pixel perturbation budget and $r \in [0,1]$ as the base ratio controlling the trade-off between uniform and focused allocation. The allocation is formulated as:
\begin{equation}
\label{eq:perturbation}
\boldsymbol{\epsilon}_{\text{map}}(i,j) = \epsilon_{\text{bg}} 
+ \frac{\mathbb{M}(i,j)}{\sum_{x=1}^{R} \sum_{y=1}^{R} \mathbb{M}(x,y)} \cdot \epsilon_{\text{fg,total}},
\end{equation}
where $\epsilon_{\text{bg}} = r \cdot \epsilon$ is a uniform base perturbation assigned to every pixel, and $\epsilon_{\text{fg,total}} = \epsilon \cdot (1 - r) \cdot R^2$ is the remaining budget for focused allocation to text-grounded regions.
This formulation ensures that the total perturbation budget across all pixels equals $\epsilon \cdot R^2$, while regions with higher grounding scores receive stronger perturbations.

\subsection{Grounding-Centric Evidence Disruption}
\label{sec:disruption}


While GPA improves transferability by guiding perturbations toward text-grounded regions, it does not ensure that these regions are perturbed with sufficient density and strength. This limitation highlights the need for explicitly reinforcing disruption in the optimization objective.
To this end, we introduce GED, a dual-faceted strategy that weakens image-text grounding at two complementary levels.  
(1) \textbf{Global disruption} induces large-scale shifts in overall visual grounding, reducing alignment with the textual description and distorting the image representation.  
(2) \textbf{Local disruption} targets region--phrase correspondences by perturbing the set of visual tokens associated with each noun phrase, ensuring that key concepts are consistently disrupted rather than only a few isolated patches.
By jointly applying these two levels, GED achieves stronger and denser perturbations that lead to a more comprehensive degradation of multimodal understanding.

\noindent\textbf{Global Disruption.}
The objective of \emph{global disruption} is twofold:  
(1) disrupting the alignment between the image and its original textual description $T_d$ in the joint vision–language embedding space; 
and (2) inducing a substantial shift in the overall grounded representation, making the adversarial image $I_{\text{adv}}$ distinct from the original clean image $I$ in grounded representation. They are realized with the following two loss terms. 


\noindent\textbf{(1) Text–Image Loss.} 
This loss measures the cosine distance between $I_{\text{adv}}$
and $T_d$ in the aligned embedding space to disrupt their global grounding alignment. It is formulated as:
\begin{equation}
\mathcal{L}_{\text{text-image}}(I_{\text{adv}}, T_d) = 1 - \cos\left(f_v(I_{\text{adv}}), f_t(T_d)\right),
\end{equation}
where $\cos(\cdot,\cdot)$ denotes the cosine similarity. By maximizing this loss, we encourage the adversarial image to become misaligned with its original textual description.

\noindent\textbf{(2) Image–Image Loss.}  
This loss measures the cosine distance between $I_{\text{adv}}$
and the original image $I$. Let $f_v^{\text{all}}(\cdot) \in \mathbb{R}^{(HW+1) \times d_v}$ denote all visual tokens from the vision encoder (including both [CLS] and $H \times W$ patch tokens) before projection.
The loss is:
\begin{equation}
\mathcal{L}_{\text{image-image}}(I_{\text{adv}}, I) = 1 - \cos\left(f_v^{\text{all}}(I_{\text{adv}}), f_v^{\text{all}}(I)\right).
\end{equation}
Using all tokens, rather than only the [CLS] embedding or the projected vision features, can capture both coarse- and fine-grained grounded visual representations, ensuring that perturbations alter detailed features and overall representation. By maximizing this loss, we distort the holistic grounded representation of $I_{\text{adv}}$.


\noindent\textbf{Local Disruption.}
Although global disruption weakens both cross-modal alignment (image--text) and unimodal consistency (image--image), it may still fail to inject sufficient density and strength into specific grounded regions, as it primarily encourages large shifts in the overall embedding space. 
To address this limitation, we introduce \emph{local disruption}, which explicitly targets visual tokens associated with each noun phrase. 
We focus on noun phrases because they provide stable visual anchors, whereas verbs often depend on the visual evidence of related objects or body parts, e.g., a person's mouth and food for ``eating'', which can already be covered by noun-grounded patches. 
We further validate this design choice in Appendix~\ref{sec:additional_ablaction_study} (Table~\ref{tab:verb_ablation}). 
By concentrating perturbations on these tokens, local disruption increases perturbation density within grounded evidence regions, thereby breaking redundant region--phrase grounding. 
We implement this process in three steps.

\noindent\textbf{Step 1: Noun Phrase Extraction.}  
Given a textual description $T_d$,  
we extract $N$ noun phrases $\{p_n\}_{n=1}^{N}$ using the SpaCy \texttt{en\_core\_web\_sm} model~\cite{spacy_en_core_web_sm},  
which detects all noun chunks.  
We remove duplicates and stop words to ensure clean phrase sets.  
Here, $N \ge 1$ in all cases due to the presence of at least one noun phrase.

\noindent\textbf{Step 2: Phrase–Token Association.}  
For each $p_n$, we compute a CLIP-based patch relevance map $\mathbf{A}_n \in \mathbb{R}^{H \times W}$ using the same Grad-ECLIP procedure from \Cref{sec:semantic_mask},  
with $p_n$ as the text input instead of $T_d$.  
Let $\mathbf{A}_n^{\text{flat}} \in \mathbb{R}^{HW \times 1}$ be its flattened version.  
The set of relevant visual token indices is:
\begin{equation}
\label{eq:R}
\mathcal{R}_n  = \{ i \in [0, HW) \mid \mathbf{A}_n^{\text{flat}}[i] > \tau \},
\end{equation}
where $\tau \in (0,1)$ is a fixed relevance threshold chosen to balance coverage and precision.  
If $\mathcal{R}_n$ is empty, the phrase is discarded from the loss computation.

\noindent\textbf{Step 3: Local Grounding Loss.}  
Let $f_v^{\text{patch}}(\cdot) \in \mathbb{R}^{HW \times d_v}$ denote the patch visual tokens (excluding [CLS]) from the final visual encoder layer.  
For each $p_n$, we compute its \emph{phrase-level visual feature} $\mathbf{c}_n \in \mathbb{R}^{d_v}$ by averaging the normalized clean visual tokens over $\mathcal{R}_n$.  
To disrupt region–phrase correspondence at a fine-grained level, we define the local grounding loss as:
\begin{equation}
\mathcal{L}_{\text{local}} (I_{\text{adv}}, \mathcal{R}_n) = \frac{1}{N} \sum_{n=1}^{N} \frac{1}{|\mathcal{R}_n|} \sum_{i \in \mathcal{R}_n} \left[ 1 - \cos\left( f_v^{\text{patch}}(I_{\text{adv}})[i], \, \mathbf{c}_n \right) \right],
\end{equation}
which penalizes similarity between adversarial visual tokens and the clean phrase-level visual feature,  
explicitly disrupting fine-grained region–phrase alignment and weakening localized multimodal understanding.

\subsection{Overall Attack Process}
Our final attack objective integrates the perturbation allocation strategy (\textit{GPA}) and the representation disruption strategy (\textit{GED}) into a unified optimization framework.  
Given a clean image $I$ and its caption $T_d$, the adversarial image is $I_{\text{adv}} = I + \delta$, where $\delta$ is constrained by a pixel-wise budget $\boldsymbol{\epsilon}_{\text{map}}$.  
The total loss combines global and local disruption:
\begin{equation}
\mathcal{L}_{\text{total}} = \mathcal{L}_{\text{text-image}} + \mathcal{L}_{\text{image-image}} + \mathcal{L}_{\text{local}}.
\end{equation}
The optimization problem is:
\begin{equation}
\label{eq:object}
\max_{\delta} \ \mathcal{L}_{\text{total}} 
\quad \text{s.t.} \quad
|\delta(x,y)|_{\infty} \leq \boldsymbol{\epsilon}_{\text{map}}(x,y), \ \forall (x,y).
\end{equation}
We solve this using PGD,  
iteratively updating $\delta$ in the gradient ascent direction and projecting it back to the $\ell_\infty$ ball defined by $\boldsymbol{\epsilon}_{\text{map}}$.  
The overall algorithm is given in \Cref{alg:sgma} in Appendix. 



%% file: Section/6_Experiment.tex
\begin{table*}[t!]
\caption{
The attack performance on open-source victim LVLMs. 
\textbf{Bold} indicates the best performance.
Results on more LVLMs and more tasks are provided in Appendix~\ref{app:full_results} (\Cref{tab:more_lvllm_results,tab:CLS_VQA}).
}
\label{tab:main_result}
\centering
\resizebox{\textwidth}{!}{
\begin{tabular}{ll|cccccc|c}
\toprule
\multirow{2}{*}{\textbf{Victim LVLM}} &\multirow{2}{*}{\textbf{Attack}} & \multicolumn{6}{c|}{\textbf{CLIP Similarity between image and generated text $\downarrow$}} & \multirow{2}{*}{\textbf{ASR (\%) $\uparrow$}}\\
&& RN-50 & RN-101 & ViT-B/16 & ViT-B/32 & ViT-L/14 & Ensemble & \\
\midrule
\multirow{9}{*}{LLaVA}
&Clean &  0.2421& 0.4646 & 0.3061& 0.2988& 0.2637&0.3151&-\\
& TGR \cite{zhangTransferable2023}&0.2364 & 0.4577 & 0.2986 & 0.2935 & 0.2527 & 0.3078& 35.6\\
& PNA \cite{wei2022towards}&0.2417 & 0.4639 & 0.3049 & 0.2998 & 0.2616 & 0.3144&22.6\\
& PNA + PathOut \cite{wei2022towards}&0.2427 & 0.4642 & 0.3063 & 0.3001 & 0.2633 & 0.3153&17.6\\
&\citet{cuiRobustnessLargeMultimodal2024}&0.2365 & 0.4584 & 0.2981 & 0.2925 & 0.2530 & 0.3077& 41.8\\
&\citet{schlarmannAdversarialRobustnessMultiModal2023} &0.2376 & 0.4600 & 0.3011 & 0.2946 & 0.2563 & 0.3099&28.4\\
&Attack-Bard~\cite{dongHowRobustGoogles2023}&0.2354 & 0.4568 & 0.2964 & 0.2915 & 0.2498 & 0.3060&38.4\\
&VT-Attack \cite{wangBreakVisualPerception2024} &0.2330 & 0.4544 & 0.2939 & 0.2892 & 0.2462 & 0.3033&46.0  \\
& \textbf{\AttackName}  &\textbf{0.2282} & \textbf{0.4493} & \textbf{0.2873} & \textbf{0.2831} & \textbf{0.2376} & \textbf{0.2971} &\textbf{55.4}  \\ 
\midrule

\multirow{9}{*}{Qwen2.5-VL}
&Clean &  0.2578 &0.4843&0.3197&0.3130&0.2702&0.3290&-\\
& TGR \cite{zhangTransferable2023}& 0.2546 & 0.4802 & 0.3149 & 0.3095 & 0.2624 & 0.3243& 24.0 \\
& PNA \cite{wei2022towards}&0.2533 & 0.4787 & 0.3131 & 0.3080 & 0.2623 & 0.3231&22.6\\
& PNA + PathOut \cite{wei2022towards}&0.2554 & 0.4810 & 0.3159 & 0.3103 & 0.2660 & 0.3257&19.0\\
&\citet{cuiRobustnessLargeMultimodal2024}&0.2530 & 0.4789 & 0.3135 & 0.3079 & 0.2620 & 0.3231&24.2\\
&\citet{schlarmannAdversarialRobustnessMultiModal2023} &0.2523 & 0.4783 & 0.3134 & 0.3075 & 0.2619 & 0.3227&26.8\\
&Attack-Bard~\cite{dongHowRobustGoogles2023}&0.2524 & 0.4788 & 0.3129 & 0.3066 & 0.2607 & 0.3223&25.0\\
&VT-Attack \cite{wangBreakVisualPerception2024} & 0.2497 & 0.4762 & 0.3105 & 0.3052 & 0.2571 & 0.3197 & 31.4 \\
& \textbf{\AttackName}  &\textbf{0.2481} & \textbf{0.4738} & \textbf{0.3070} & \textbf{0.3019} & \textbf{0.2540} & \textbf{0.3169}& \textbf{39.0} \\ 
\midrule

\multirow{9}{*}{InternVL3}
&Clean &  0.2595&0.4870&0.3240&0.3155&0.2791&0.3330&-\\
& TGR \cite{zhangTransferable2023}& 0.2550 & 0.4809 & 0.3166 & 0.3100 & 0.2683 & 0.3262& 22.0\\
& PNA \cite{wei2022towards}&0.2567 & 0.4825 & 0.3184 & 0.3129 & 0.2717 & 0.3284&18.8\\
& PNA + PathOut \cite{wei2022towards}&0.2589 & 0.4841 & 0.3213 & 0.3159 & 0.2752 & 0.3311&13.4\\
&\citet{cuiRobustnessLargeMultimodal2024}&0.2534 & 0.4803 & 0.3159 & 0.3099 & 0.2682 & 0.3255&24.6\\
&\citet{schlarmannAdversarialRobustnessMultiModal2023} &0.2552 & 0.4808 & 0.3168 & 0.3098 & 0.2695 & 0.3264&23.0\\
&Attack-Bard~\cite{dongHowRobustGoogles2023}&0.2535 & 0.4798 & 0.3146 & 0.3087 & 0.2663 & 0.3246&24.2\\
&VT-Attack \cite{wangBreakVisualPerception2024} &0.2509 & 0.4776 & 0.3125 & 0.3066 & 0.2632 & 0.3222&  31.6\\
& \textbf{\AttackName}  & \textbf{0.2474} & \textbf{0.4735} & \textbf{0.3072} & \textbf{0.3026} & \textbf{0.2574} & \textbf{0.3176} & \textbf{41.2} \\ 
\bottomrule
\end{tabular}}
\end{table*}

\section{Evaluation}
\label{sec:experiment}

\subsection{Experimental Setup}
\label{sec:experimental_setup}

\textbf{Surrogate and Victim Models.}  
We employ the same set of victim models as detailed in \Cref{sec:additional_setup}. 
For surrogate models, we select CLIP-L/14 (default), CLIP-B/16, SigLIP~\cite{zhai2023sigmoid} and DINOv2-B~\cite{oquab2023dinov2} to investigate the impact of surrogates.

\noindent\textbf{Datasets.}  
We randomly sample 1,000 samples from each of three datasets for evaluation, corresponding to different multimodal tasks: image captioning on Flickr30k~\cite{young2014image} using the prompt ``Describe the image in one sentence'', image classification on CIFAR-10~\cite{krizhevsky2009learning} using the prompt in \Cref{sec:additional_setup}, and visual question answering (VQA) on the TextVQA validation set~\cite{singh2019towards}.
Diverse datasets cover both object-centric recognition and text-based reasoning scenarios.

\noindent\textbf{Metrics.}  
We use attack success rate (ASR) to quantify attack effectiveness across all tasks, though the computation method varies depending on the task nature. For classification and VQA, ASR is the fraction of adversarial samples that change a correct clean prediction into an incorrect one. 
For captioning, where no ground-truth ``correct/incorrect'' label is available, we follow~\cite{xieChainAttackRobustness2025} and use an \textit{LVLM-as-a-Judge} protocol to assess attack success (see Appendix~\ref{sec:additional_setup}). 
Following~\cite{zhaoEvaluatingAdversarialRobustness2023, wangBreakVisualPerception2024}, we also report CLIP similarity between the adversarially generated text $y_{\text{adv}}$ and the clean image $I$ for captioning, where lower CLIP similarity and higher ASR indicate stronger attacks.

\noindent\textbf{Baselines.}  
We compare our proposed \AttackName with four representative LVLM attacks~\cite{cuiRobustnessLargeMultimodal2024, schlarmannAdversarialRobustnessMultiModal2023, dongHowRobustGoogles2023, wangBreakVisualPerception2024}, 
as well as three transfer-based attacks originally designed for ViTs~\cite{wei2022towards, zhangTransferable2023}. 
Specifically, TGR~\cite{zhangTransferable2023} regularizes token gradients to reduce variance and concentrate perturbations on more transferable components, 
while PNA and PatchOut~\cite{wei2022towards} improve transferability by treating attention weights as constants and randomly masking patches during backpropagation. 
We adapt these transfer-based methods to the LVLM setting by integrating them into the framework of \citet{cuiRobustnessLargeMultimodal2024}.


\noindent\textbf{Attack Setting.}  
For each image, we obtain a concise description $T_d$ from GPT-4o using the prompt ``Describe this image in a short sentence.''  
Perturbations are bounded by $\epsilon = 8/255$ under the $\ell_\infty$ norm and optimized with PGD using $K = 100$ steps and step size $\alpha = 1/255$.  
We set the default values of base ratio $r = 0.2$ and relevance threshold $\tau = 0.3$. 
All experiments are performed on a cluster equipped with NVIDIA GeForce RTX 4090 GPUs.

\subsection{Experimental Results}

\noindent\textbf{Main Results.} We evaluate the effectiveness of \AttackName across eight popular LVLMs and compare it against existing adversarial attacks, with the main results shown in \Cref{tab:main_result}. Furthermore, we assess the imperceptibility of generated adversarial examples in Appendix~\ref{sec:attack_imperceptiblity}, and provide visual illustrations in Appendix~\ref{sec:visual_illustration} to explain why \AttackName achieves higher transferability. Overall, \AttackName consistently achieves lower image--text CLIP similarity and higher ASR than prior encoder-based attacks, demonstrating strong transferability across diverse open-source LVLMs, including LLaVA, Qwen2.5-VL, InternVL3, OpenFlamingo, BLIP-2, and Kimi-VL. The gains are especially clear on heterogeneous victims such as Qwen2.5-VL and InternVL3, which suggests that grounding-driven localization improves transfer beyond architectures closely aligned with the surrogate. At the same time, we find that transfer-enhancing strategies originally designed for ViTs, such as PNA~\cite{wei2022towards}, PatchOut~\cite{wei2022towards}, and TGR~\cite{zhangTransferable2023}, remain limited in LVLMs. A likely reason is that these methods mainly perturb the CLS token, which is effective for classification models relying on a global representation, but less effective for LVLMs that depend on fine-grained patch-level evidence. PatchOut is particularly unfavorable in this setting, since discarding patches reduces perturbation density on key grounded regions instead of disrupting the redundant visual tokens multimodal generation relies on.

\begin{table*}[t!]
\caption{
Attack performance on commercial victim LVLMs when all methods are equipped with both model ensemble (ME) and diverse input (DI).
\textbf{Bold} indicates the best performance.
}
\label{tab:ME/DI_performance}
\centering
\setlength{\tabcolsep}{4pt}
\renewcommand{\arraystretch}{0.95}
\resizebox{0.9\textwidth}{!}{
\begin{tabular}{ll|cccccc|c}
\toprule
\multirow{2}{*}{\textbf{Victim LVLM}} & \multirow{2}{*}{\textbf{Attack}} & \multicolumn{6}{c|}{\textbf{CLIP Similarity between image and generated text $\downarrow$}} & \multirow{2}{*}{\textbf{ASR (\%) $\uparrow$}}\\
&& RN-50 & RN-101 & ViT-B/16 & ViT-B/32 & ViT-L/14 & Ensemble & \\
\midrule
\multirow{4}{*}{GPT-4o}
&\citet{cuiRobustnessLargeMultimodal2024} & 0.2226 & 0.4481 & 0.2765 & 0.2728 & 0.2198 & 0.2880 & 52.0 \\
&Attack-Bard~\cite{dongHowRobustGoogles2023} & 0.2280 & 0.4540 & 0.2837 & 0.2796 & 0.2316 & 0.2954 & 45.7 \\
&VT-Attack~\cite{wangBreakVisualPerception2024}& 0.2151 & 0.4408 & 0.2702 & 0.2668 & 0.2132 & 0.2812 & 64.3 \\
&\textbf{\AttackName}  & \textbf{0.1943} & \textbf{0.4198} & \textbf{0.2439} & \textbf{0.2394} & \textbf{0.1876} & \textbf{0.2570} & \textbf{80.0} \\
\midrule
\multirow{4}{*}{Gemini 2.0 Flash}
&\citet{cuiRobustnessLargeMultimodal2024} & 0.2430 & 0.4695 & 0.3019 & 0.2935 & 0.2495 & 0.3115 & 45.6 \\
&Attack-Bard~\cite{dongHowRobustGoogles2023} & 0.2415 & 0.4677 & 0.2980 & 0.2927 & 0.2490 & 0.3098 & 36.9 \\
&VT-Attack~\cite{wangBreakVisualPerception2024}& 0.2289 & 0.4556 & 0.2839 & 0.2812 & 0.2317 & 0.2963 & 58.5 \\
&\textbf{\AttackName}  & \textbf{0.2173} & \textbf{0.4440} & \textbf{0.2682} & \textbf{0.2663} & \textbf{0.2173} & \textbf{0.2826} & \textbf{66.8} \\
\midrule
\multirow{4}{*}{GPT-5.4}
&\citet{cuiRobustnessLargeMultimodal2024} & 0.2345 & 0.4579 & 0.2846 & 0.2854 & 0.2279 & 0.2981 & 59.0 \\
&Attack-Bard~\cite{dongHowRobustGoogles2023} & 0.2333 & 0.4576 & 0.2843 & 0.2836 & 0.2307 & 0.2979 & 53.3 \\
&VT-Attack~\cite{wangBreakVisualPerception2024} & 0.2271 & 0.4518 & 0.2759 & 0.2776 & 0.2197 & 0.2904 & 63.7 \\
&\textbf{\AttackName} & \textbf{0.2078} & \textbf{0.4321} & \textbf{0.2549} & \textbf{0.2545} & \textbf{0.1953} & \textbf{0.2689} & \textbf{81.7} \\
\bottomrule
\end{tabular}}
\vspace{-2mm}
\end{table*}

\noindent\textbf{Attack Performance on Commercial LVLMs.}
Closed-source commercial LVLMs present a more realistic but challenging black-box setting, where direct transfer attacks under the default setting are less effective due to the larger surrogate-victim gap. 
Therefore, for commercial victims, we report the results under a stronger setting that combines diverse input (DI) and model ensemble (ME) in Table~\ref{tab:ME/DI_performance}. 
Here, DI improves robustness to input transformations, while ME reduces surrogate-specific overfitting by optimizing across three CLIP surrogates, namely ViT-L/14, ViT-B/16, and ViT-B/32. This setting provides a stronger and more practical evaluation for attacking commercial LVLMs.
As shown in Table~\ref{tab:ME/DI_performance}, GDA consistently improves attack performance across all three commercial LVLMs, demonstrating that our method remains effective beyond open-source victims. 
Moreover, DI and ME enhance transferability through input-level diversity and model/ensemble-level enhancement, while GDA improves the attack objective by explicitly disrupting text-grounded visual evidence. 
Thus, GDA is orthogonal to these generic transfer boosters and can be naturally combined with them. 
The consistent gains under the DI+ME setting further show that our grounding-driven design can be integrated with existing black-box transfer-enhancement strategies to strengthen attacks on commercial LVLMs.

Due to space limitations, we include additional analyses in Appendix~\ref{sec:additional_ablaction_study}, covering component ablation, surrogate models, perturbation budget $\epsilon$, base ratio $r$, relevance threshold $\tau$, the number of attack steps, computational cost, adding verbs to the local grounding term, and caption quality in Step~1. We further discuss potential defenses and limitations in Appendix~\ref{sec:discussion}.


%% file: Section/Appendix.tex
\newpage
\appendix

\section{Appendix Overview}
\label{sec:appendix_summary}
We briefly outline the appendix structure. 
\Cref{sec:broader_impacts_appendix} discusses broader impacts. \Cref{sec:notation} lists the key notation, and \Cref{sec:related_work} reviews related work. \Cref{sec:case_study} presents additional attack scenarios and a medical-diagnosis case study. \Cref{sec:alg} provides pseudocode for \AttackName, and \Cref{sec:visual_illustration} presents qualitative visual illustrations. \Cref{sec:additional_setup} details the experimental setup, while \Cref{sec:additional_exmperimental_results} reports additional experimental results. Finally, \Cref{sec:discussion} discusses possible defenses and limitations.


\section{Broader Impacts}
\label{sec:broader_impacts_appendix}
This work has both positive and negative societal implications. On the positive side, it improves our understanding of transferable vulnerabilities in LVLMs and can support more realistic robustness evaluation, stronger defenses, and safer deployment practices for multimodal systems in safety-critical domains. On the negative side, the proposed attack could be misused to manipulate model outputs in applications such as content moderation, medical decision support, or autonomous perception. We therefore present the method in the context of robustness analysis, discuss defenses and limitations, and do not release any high-risk new dataset or model asset as part of this work.

\section{Notation}
\label{sec:notation}
We provide the frequently used notation throughout this paper for reference in \Cref{tab:notation}.

\begin{table}[h]
\centering
\caption{Notation used throughout the paper.}
\begin{tabular}{cl}
\toprule
\textbf{Symbol} & \textbf{Description} \\
\midrule
$I$ & Input image \\
$T_d$ & Text description corresponding to the $I$\\
$I_{\text{adv}}$ & Adversarial image ($I + \delta$) \\
$\mathbf{v}, \mathbf{t}$ & Normalized image and text embeddings \\
$d$ & Shared alignment dimension (after projection) \\
$d_v$ & Vision encoder hidden dimension (before projection) \\
$R$ & Input image resolution \\
$H \times W$ & Patch grid size \\
$\mathbf{A}$ & Patch-level attention map \\
$\mathbb{M}$ & Pixel-level grounding mask \\
$\boldsymbol{\epsilon}_{\text{map}}$ & Pixel-wise perturbation map \\
$\epsilon$ & Average per-pixel perturbation budget \\
$r$ & Base ratio for uniform vs. focused allocation \\
$f_v(\cdot)$ & CLIP vision encoder \\
$f_t(\cdot)$ & CLIP text encoder \\
$f_v^{\text{all}}(\cdot)$ & All visual tokens from vision encoder (before projection) \\
$f_v^{\text{patch}}(\cdot)$ & Patch visual tokens from vision encoder (excluding [CLS]) \\
\bottomrule
\end{tabular}
\label{tab:notation}
\end{table}

\section{Related Work}
\label{sec:related_work}
\input{Section/Related_work}

\section{Attack Scenarios}
\label{sec:case_study}

The practical feasibility of adversarial attacks poses serious risks when LVLMs are deployed in safety- and security-critical domains such as medical image analysis~\cite{hu2024omnimedvqa,wu2024pmc,rao2025multimodal,fahrner2025generative} and autonomous driving perception systems~\cite{chen2025automated}.
In medical applications, an adversary could introduce imperceptible perturbations at various points in the imaging pipeline, e.g., via compromised scanners, storage systems, or network infrastructure~\cite{mirsky2019ct}. 
Such perturbations may cause LVLMs to produce misleading diagnostic outputs, potentially leading to erroneous clinical decisions. 
In autonomous driving, perturbed images could alter the model’s interpretation of traffic signs or critical objects, leading to unsafe control decisions or failures in threat detection \cite{10.1145/3691625}. 
These scenarios illustrate the real-world risks posed by transferable adversarial examples against LVLMs, which we further substantiate through a case study on medical image diagnosis.

\noindent\textbf{Case Study on Medical Diagnostic.}
\begin{figure}[t]
    \centering
    \includegraphics[width=0.82\linewidth]{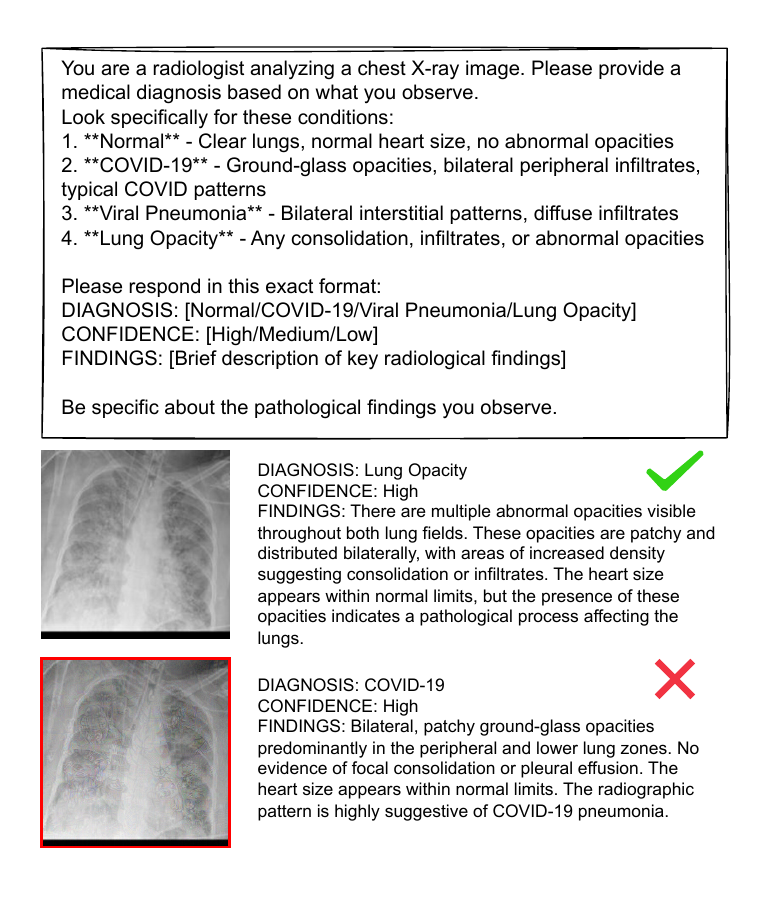}
    \caption{Illustration of diagnostic errors on chest X-ray images generated by \AttackName. The top image is correctly diagnosed as Lung Opacity, while the bottom image is misclassified as COVID-19 with high confidence.}
    \label{fig:case_study}
\end{figure}
We conduct a case study on the COVID-19 Radiography dataset~\cite{rahman2021covid19radiography} to demonstrate the \emph{real-world risks} of adversarial attacks in safety-critical domains. 
This case mimics a typical \emph{medical diagnostic setting}, where chest X-ray images must be classified into clinically relevant categories to guide radiologists in disease identification and treatment decisions. 
We construct the evaluation set by sampling 100 chest X-ray images evenly across four diagnostic categories: Normal, COVID-19, Viral Pneumonia, and Lung Opacity. 
On this set, we apply our proposed attack, \AttackName, to generate adversarial counterparts under the same configuration described in \Cref{sec:experimental_setup}, and we quantify its effectiveness using the ASR, where a diagnostic misclassification is regarded as a successful attack. 
We evaluate the impact of these adversarial examples on GPT-4.1, treated as the victim model, with a structured radiology-style diagnostic prompt shown in \Cref{fig:case_study}.

\noindent\textbf{Results.} \AttackName achieves an ASR of \textbf{61\%}, highlighting the alarming susceptibility of LVLM-based diagnostic systems to adversarial manipulation in realistic clinical settings. Beyond the aggregate number, the adversarial perturbations lead to particularly concerning high-stakes errors. Figure~\ref{fig:case_study} illustrates one such case: on the clean image, GPT-4.1 outputs ``\texttt{DIAGNOSIS: Lung Opacity, CONFIDENCE: High}'', accompanied by findings describing multiple abnormal opacities and bilateral consolidations. Under adversarial perturbation, however, the same image is misclassified as ``\texttt{DIAGNOSIS: COVID-19, CONFIDENCE: High}'', with fabricated findings such as “bilateral, patchy ground-glass opacities predominantly in the peripheral and lower lung zones.” From a clinical perspective, lung opacity and COVID-19 are already challenging to distinguish, and such adversarially induced misclassification, presented with high confidence and detailed but spurious radiological descriptions, can be especially misleading for medical decision-making.


\noindent\textbf{Implications.} These results demonstrate that even imperceptible perturbations can induce systematic and confident misdiagnoses in medical scenarios. Such failures pose direct threats to clinical decision-making, as adversarially manipulated images could mislead radiologists or automated triage systems, ultimately endangering patient safety. In practice, an adversary could embed perturbations at multiple stages of the imaging pipeline, for instance by compromising X-ray scanners, tampering with picture archiving and communication systems (PACS) used in hospitals, or injecting malicious noise during cloud-based storage and transmission~\cite{mirsky2019ct}. Once introduced, these perturbations may cause LVLMs to output misleading diagnoses that could result in inappropriate treatments (e.g., unnecessary isolation, antiviral medication) or delayed care for the true underlying condition. 

This case study highlights that adversarial attacks represent not only an academic concern but a tangible threat to medical AI deployments. The ability to covertly alter diagnostic outputs without raising visual suspicion underscores the urgent need for robust defenses, including adversarial training, input verification, and end-to-end security auditing, before deploying LVLMs in safety- and security-critical applications.

\section{Pseudocode of \AttackName}
\label{sec:alg}
We present the detailed pseudocode of \AttackName in~\Cref{alg:sgma}.

\begin{algorithm}[t]
\caption{Grounding-Driven Attack (\AttackName)}
\label{alg:sgma}
\begin{algorithmic}[1]
\REQUIRE Clean image $I$, steps $K$, step size $\alpha$, perturbation budget $\boldsymbol{\epsilon}$, base ratio $r$, relevance  threshold $\tau$
\ENSURE Adversarial image $I_{\text{adv}}$

\STATE \textcolor{blue}{\textit{\# Step 1: Grounding-Aware Perturbation Allocation (GPA)}}
\STATE Generate description caption $T_d$ of image $I$ via a proxy LVLM.
\STATE Compute the grounding mask $\mathbb{M}$
\STATE Generate the perturbation allocation map $\boldsymbol{\epsilon}_{\text{map}}$ using \Cref{eq:perturbation} with base ratio $r$

\STATE \textcolor{blue}{\textit{\# Step 2: Grounding-Centric Evidence Disruption (GED)}}
\STATE Extract noun phrases $\{p_n\}_{n=1}^N$ from $T_d$
\FOR{each phrase $p_n$}
    \STATE Identify aligned patch indices $\mathcal{R}_n$ based on relevance threshold $\tau$
    \STATE Compute phrase-specific center $\mathbf{c}_n$ from clean patch features over $\mathcal{R}_n$
\ENDFOR
\STATE Initialize perturbation $\delta \gets 0$ and adversarial image $I_{\text{adv}} \gets I + \delta$
\FOR{$t = 1$ to $K$}
    \STATE Compute total loss: $\mathcal{L}_{\text{total}} = \mathcal{L}_{\text{text-image}} + \mathcal{L}_{\text{image-image}} + \mathcal{L}_{\text{local}}$
    \STATE Gradient ascent: $\delta \gets \delta + \alpha \cdot \text{sign}\left(\nabla_\delta \mathcal{L}_{\text{total}}\right)$
    \STATE Budget clipping: $\delta \gets \text{clip}(\delta, -\boldsymbol{\epsilon}_{\text{map}}, \boldsymbol{\epsilon}_{\text{map}})$
    \STATE Update adversarial image: $I_{\text{adv}} \gets I + \delta$
\ENDFOR
\RETURN $I_{\text{adv}}$
\end{algorithmic}
\end{algorithm}


\section{Visual Illustration}
\label{sec:visual_illustration}
\begin{table}[t]
\centering
\caption{
Grad-ECLIP-based perturbation--saliency alignment comparison between VT and GDA (mean $\pm$ std, $N=100$).
}
\setlength{\tabcolsep}{6pt}
\begin{tabular}{lcc}
\toprule
\textbf{Metric (Grad-ECLIP-based)} & \textbf{VT-Attack} & \textbf{GDA} \\
\midrule
$\mathrm{Spearman}(\mathrm{pert},\mathrm{sal})$ $\uparrow$ & $0.0524 \pm 0.0834$ & $\mathbf{0.5994 \pm 0.0832}$ \\
$\mathrm{IoU@Top20\%}(\mathrm{pert},\mathrm{sal})$ $\uparrow$ & $0.1303 \pm 0.0443$ & $\mathbf{0.2936 \pm 0.0709}$ \\
\midrule
$\mathrm{coverage@Top20\%\_sal}$ $\uparrow$ & $0.4108 \pm 0.0868$ & $\mathbf{0.5196 \pm 0.1019}$ \\
\bottomrule
\end{tabular}
\label{tab:grad_eclip_vt_vs_sgma}
\end{table}

\begin{figure*}
    \centering
    \includegraphics[width=\linewidth]{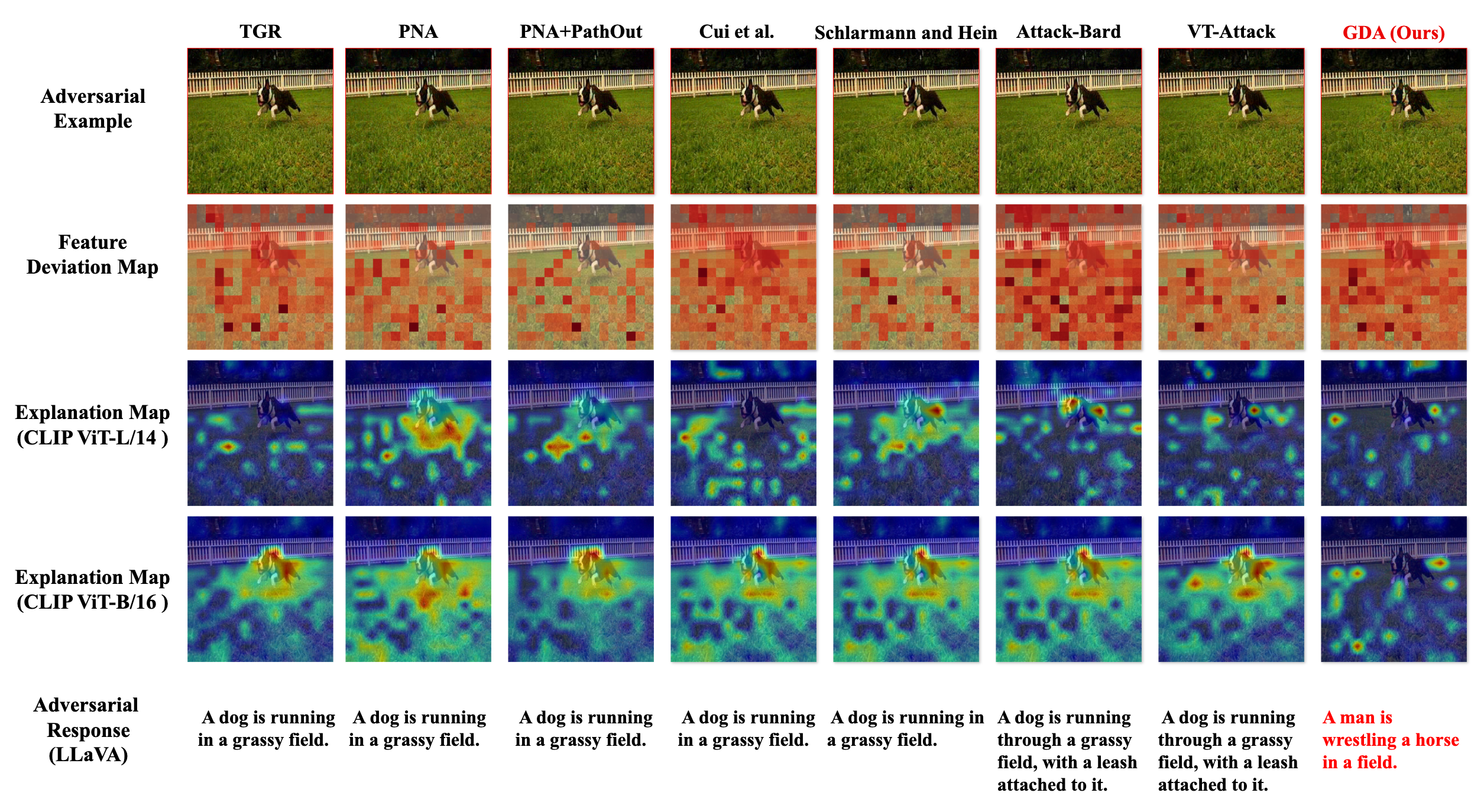}
  \caption{Qualitative comparison of adversarial effectiveness across different attack methods.}
    \label{Fig:perturbation_comparison}
\end{figure*}

\begin{figure*}
    \centering
    \includegraphics[width=1\linewidth]{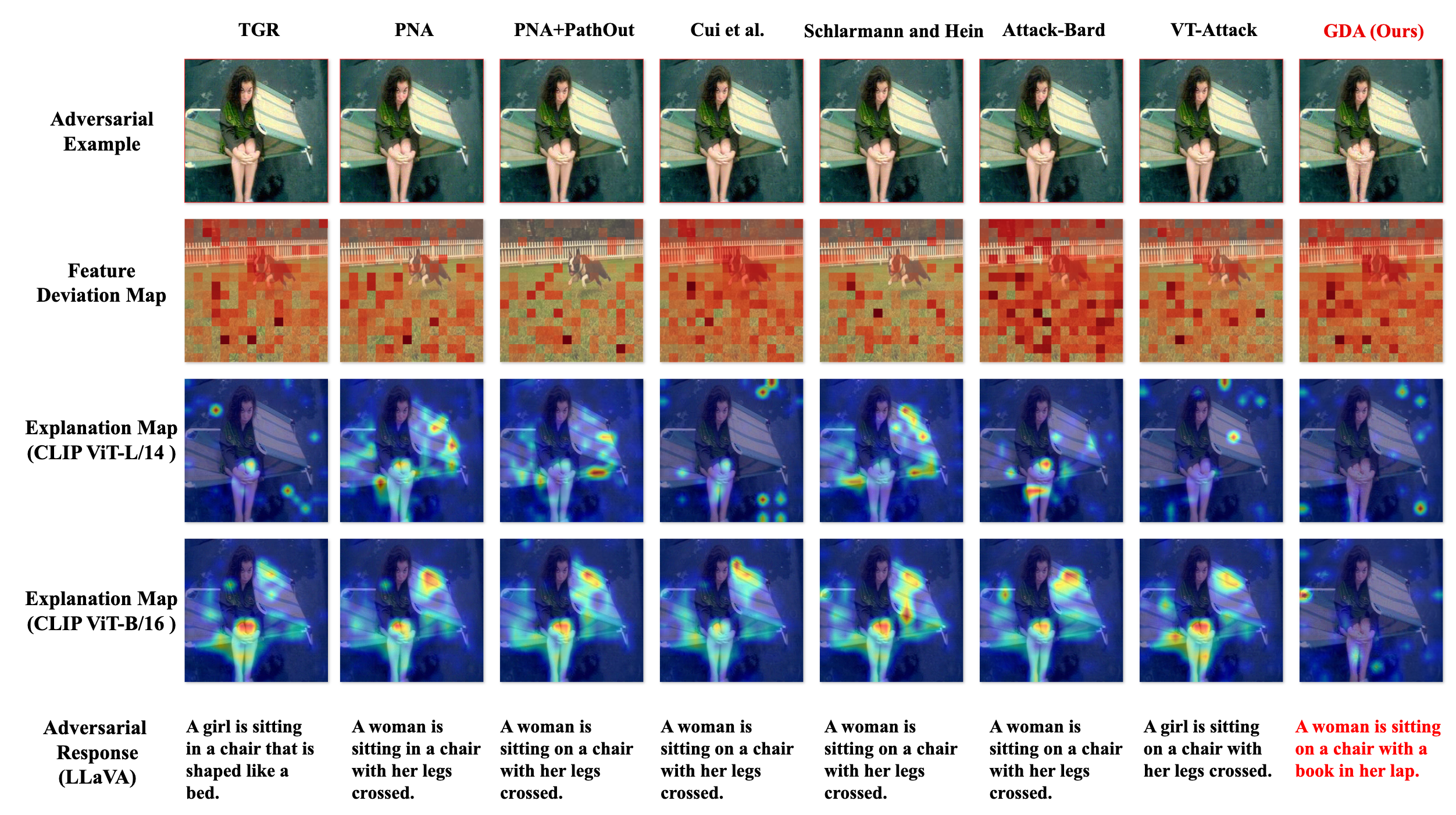}
    \caption{Qualitative comparison of adversarial effectiveness across different attack methods.}
    \label{fig:example_appendix1}
\end{figure*}

To understand \AttackName's superior performance, we compare adversarial examples in \Cref{Fig:perturbation_comparison,fig:example_appendix1} from three complementary views:
(1) feature deviation maps, showing patch-wise changes in CLIP ViT-L/14 visual embeddings relative to the clean image;
(2) explanation maps computed using Grad-ECLIP~\cite{zhaoGradientbasedVisualExplanation2024} with the same CLIP ViT-L/14 encoder used as the surrogate during attack generation; and
(3) explanation maps computed using Grad-ECLIP~\cite{zhaoGradientbasedVisualExplanation2024} with a different CLIP ViT-B/16 encoder, allowing us to examine whether perturbation effects persist across encoders that differ from the surrogate.
Table~\ref{tab:grad_eclip_vt_vs_sgma} first provides a quantitative comparison between VT-Attack and \AttackName. Relative to VT-Attack, \AttackName achieves substantially stronger perturbation--saliency alignment under all three Grad-ECLIP-based metrics, improving Spearman correlation from $0.0524$ to $0.5994$, IoU@Top20\% from $0.1303$ to $0.2936$, and coverage@Top20\%\_sal from $0.4108$ to $0.5196$. These gains indicate that \AttackName not only concentrates perturbations more effectively on grounded evidence regions, but also covers a larger fraction of the patches identified by Grad-ECLIP as most relevant to the text.

This quantitative trend is clearly reflected in the qualitative examples. As shown in \Cref{Fig:perturbation_comparison}, methods such as \citet{schlarmannAdversarialRobustnessMultiModal2023} and VT-Attack~\cite{wangBreakVisualPerception2024} make weak, scattered changes that leave critical foreground objects (e.g., the dog) largely intact. Others, such as \citet{cuiRobustnessLargeMultimodal2024} and Attack-Bard~\cite{dongHowRobustGoogles2023}, perturb the object more strongly but also waste budget on irrelevant background areas such as grass, limiting their impact on grounding alignment. Notably, for these methods, the explanation maps on the surrogate encoder often show some shift, but this effect largely disappears when visualized with a different encoder, where attention remains centered on the main object.
In contrast, \AttackName focuses perturbations on text-grounded evidence regions, avoids unnecessary background changes, and produces substantial feature deviations in key areas. Even under a different encoder from the surrogate, \AttackName's explanation maps still exhibit pronounced shifts away from the correct object (e.g., ``dog''), indicating a stronger and more persistent disruption of cross-modal grounding. These qualitative examples are consistent with the higher alignment and coverage scores in Table~\ref{tab:grad_eclip_vt_vs_sgma}.

\begin{table}
    \centering
    \caption{Victim Models}
    \begin{tabular}{l|p{6cm}}
    \toprule
         \textbf{Model} & \textbf{Link} \\ \midrule
         Blip-2 & \href{https://huggingface.co/Salesforce/blip2-opt-2.7b}{Salesforce/blip2-opt-2.7b} \\
         LLaVA & \href{https://huggingface.co/llava-hf/llava-1.5-7b-hf}{llava-hf/llava-1.5-7b-hf} \\
         Qwen2.5-VL & \href{https://huggingface.co/Qwen/Qwen2.5-7B-Instruct}{Qwen/Qwen2.5-7B-Instruct} \\
         InternVL3 & \href{https://huggingface.co/OpenGVLab/InternVL3-8B}{OpenGVLab/InternVL3-8B} \\
         OpenFlamingo & \href{https://huggingface.co/openflamingo/OpenFlamingo-4B-vitl-rpj3b-langinstruct}{OpenFlamingo-4B-vitl-rpj3b-langinstruct} \\
         Kimi-VL & \href{https://huggingface.co/moonshotai/Kimi-VL-A3B-Instruct}{Kimi-VL-A3B-Instruct} \\
         GPT-4o & \href{https://platform.openai.com/docs/models/gpt-4o}{gpt-4o documentation} \\
         Gemini & \href{https://cloud.google.com/vertex-ai/generative-ai/docs/models/gemini/2-0-flash}{Gemini 2.0 Flash} \\
         GPT-5.4 & \href{https://developers.openai.com/api/docs/models/gpt-5.4}{GPT-5.4}\\
    \bottomrule
    \end{tabular}
    \label{tab:victim_model}
\end{table}

\begin{table*}[t!]
\caption{The attack performance on more LVLMs. \textbf{Bold} indicates the best performance (lowest similarity or highest ASR). The end-to-end attack \cite{schlarmannAdversarialRobustnessMultiModal2023} performs particularly well on BLIP-2, as the adversarial examples are generated directly on BLIP-2 in a white-box setting.}
\label{tab:more_lvllm_results}
\centering
\resizebox{\textwidth}{!}{
\begin{tabular}{ll|cccccc|c}
\toprule
\multirow{2}{*}{\textbf{Victim LVLM}} &\multirow{2}{*}{\textbf{Attack}} & \multicolumn{6}{c|}{\textbf{CLIP Similarity between image and generated text $\downarrow$}} & \multirow{2}{*}{\textbf{ASR (\%) $\uparrow$}}\\
&& RN-50 & RN-101 & ViT-B/16 & ViT-B/32 & ViT-L/14 & Ensemble & \\
\midrule

\multirow{9}{*}{OpenFlamingo}
&Clean & 0.2431& 0.4715& 0.3096&0.3009&0.2680&0.3186&-\\
& TGR \cite{zhangTransferable2023}& 0.1615 & 0.3878 & 0.2116 & 0.2100 & 0.1501 & 0.2242& 95.0\\
& PNA \cite{wei2022towards}&0.2118 & 0.4367 & 0.2672 & 0.2654 & 0.2145 & 0.2791&53.4\\
& PNA + PathOut \cite{wei2022towards}&0.2318 & 0.4577 & 0.2911 & 0.2880 & 0.2430 & 0.3023&70.9\\
&\citet{cuiRobustnessLargeMultimodal2024}&0.1598 & 0.3850 & 0.2108 & 0.2098 & 0.1501 & 0.2231&97.0\\
&\citet{schlarmannAdversarialRobustnessMultiModal2023} &0.2272 & 0.4544 & 0.2887 & 0.2835 & 0.2411 & 0.2990&59.4\\
&Attack-Bard~\cite{dongHowRobustGoogles2023}&0.1469 & 0.3736 & 0.1941 & 0.1959 & 0.1274 & 0.2076&\textbf{97.8}\\
&VT-Attack \cite{wangBreakVisualPerception2024} & 0.1297 & 0.3568 & 0.1785 & 0.1799 & 0.1165 & 0.1923 &  \textbf{99.8}\\
& \textbf{\AttackName}  &  \textbf{0.1263} & \textbf{0.3559} & \textbf{0.1765} & \textbf{0.1778}& \textbf{0.1129} & \textbf{0.1899}& \textbf{99.8} \\ 
\midrule

\multirow{9}{*}{BLIP-2}
&Clean &0.2402 & 0.4689 & 0.3038 & 0.2984 & 0.2585 & 0.3140 &-\\
& TGR \cite{zhangTransferable2023}& 0.2354 & 0.4623 & 0.2961 & 0.2933 & 0.2499 & 0.3074&23.4 \\
& PNA \cite{wei2022towards}&0.2365 & 0.4648 & 0.2980 & 0.2943 & 0.2543 & 0.3096&21.9\\
& PNA + PathOut \cite{wei2022towards}&0.2372 & 0.4650 & 0.2989 & 0.2952 & 0.2545 & 0.3102&17.6\\
&\citet{cuiRobustnessLargeMultimodal2024}&0.2335 & 0.4606 & 0.2957 & 0.2913 & 0.2478 & 0.3058&26.8\\
&\citet{schlarmannAdversarialRobustnessMultiModal2023} &\textbf{0.1795} & \textbf{0.4087} &\textbf{0.2339} & \textbf{0.2338} & \textbf{0.1758} & \textbf{0.2463}&\textbf{84.4}\\
&Attack-Bard~\cite{dongHowRobustGoogles2023}&0.2330 & 0.4606 & 0.2941 & 0.2900 & 0.2455 & 0.3047&26.0\\
&VT-Attack \cite{wangBreakVisualPerception2024} &  0.2286 & 0.4567 & 0.2907 & 0.2862 & 0.2409 & 0.3006&27.6  \\
& \textbf{\AttackName}  & 0.2256 & 0.4536 & 0.2857 & 0.2833 & 0.2356 & 0.2968& 38.6 \\ 
\midrule

\multirow{9}{*}{Kimi-VL}
&Clean &  0.2438&0.4654&0.3064&0.3012&0.2612&0.3156&-\\
& TGR \cite{zhangTransferable2023}&0.2411 & 0.4624 & 0.3004 & 0.2978 & 0.2555 & 0.3115&19.2 \\
& PNA \cite{wei2022towards}&0.2403 & 0.4613 & 0.3005 & 0.2981 & 0.2542 & 0.3109&17.4\\
& PNA + PathOut \cite{wei2022towards}&0.2409 & 0.4629 & 0.3020 & 0.2994 & 0.2560 & 0.3123&20.6\\
&\citet{cuiRobustnessLargeMultimodal2024}&0.2383 & 0.4592 & 0.3001 & 0.2956 & 0.2518 & 0.3090&21.8\\
&\citet{schlarmannAdversarialRobustnessMultiModal2023} &0.2382 & 0.4595 & 0.3002 & 0.2958 & 0.2512 & 0.3090&23.2\\
&Attack-Bard~\cite{dongHowRobustGoogles2023}&0.2378 & 0.4588 & 0.2993 & 0.2950 & 0.2510 & 0.3084&23.0\\
&VT-Attack \cite{wangBreakVisualPerception2024} &0.2383 & 0.4586 & 0.2986 & 0.2947 & 0.2505 & 0.3082 & 28.2 \\
& \textbf{\AttackName}  & \textbf{0.2347} & \textbf{0.4546} & \textbf{0.2943} & \textbf{0.2914} & \textbf{0.2438} & \textbf{0.3037} & \textbf{33.2}  \\ 
\bottomrule
\end{tabular}}
\end{table*}

\begin{table*}[t!]
\caption{ASR (\%) on image classification task and VQA task.}
\label{tab:CLS_VQA}
\centering
\resizebox{\textwidth}{!}{
\begin{tabular}{l|cccccccc|c}
\toprule
\textbf{Attack} & \textbf{LLaVA} & \textbf{Qwen2.5-VL} & \textbf{InternVL3} & \textbf{OpenFlamingo} & \textbf{BLIP-2} & \textbf{Kimi-VL} & \textbf{GPT-4o} & \textbf{Gemini 2.0} & \textbf{Average} \\
\midrule

\multicolumn{10}{c}{\multirow{1}{*}{\textit{\textbf{Task: Image Classification}}}}\\ \midrule
Clean & 7.2 & 9.6 & 17.2 & 54.0 & 28.4 & 13.4 & 2.2 & 8.0 & 17.5 \\
TGR \cite{zhangTransferable2023}&49.2 & 42.4 & 58.4 & 98.2 & 69.8 & 45.0 & 27.4 & 24.6 & 51.9 \\
PNA \cite{wei2022towards}&18.8 & 25.0 & 35.8 & 91.8 & 40.2 & 23.8 & 9.6 & 16.6 & 32.7 \\
PNA + PathOut \cite{wei2022towards}&17.6 & 23.4 & 30.6 & 78.6 & 38.2 & 22.8 & 11.4 & 16.0 & 29.4 \\
\citet{cuiRobustnessLargeMultimodal2024} & 47.4 & 38.2 & 55.2 & 98.4 & 68.0 & 42.2 & 26.2 & 23.2 & 50.0 \\
\citet{schlarmannAdversarialRobustnessMultiModal2023} & 21.2 & 28.4 & 45.4 & 71.6 & 82.8 & 28.0 & 12.4 & 15.4 & 38.2 \\
Attack-Bard~\cite{dongHowRobustGoogles2023} & 29.6 & 38.8 & 51.8 & 79.4 & 55.0 & 50.8 & 21.0 & 22.6 & 43.4 \\
VT-Attack~\cite{wangBreakVisualPerception2024} & 64.2 & 47.0 & 61.8 & 100.0 & 82.8 & 53.0 & 27.2 & 27.4 & 57.4 \\
\textbf{\AttackName} & 65.2 & 51.0 & 67.4 & 99.6 & 84.0 & 54.6 & 32.0 & 31.8 & \textbf{60.7} \\
\midrule

\multicolumn{10}{c}{\multirow{1}{*}{\textit{\textbf{Task: VQA}}}}\\ \midrule
Clean & 51.2 & 19.4 & 41.6 & 90.0 & 88.8 & 8.0 & 19.4 & 17.2 & 41.4 \\
TGR \cite{zhangTransferable2023} & 59.2 & 53.6 & 56.0 & 97.6 & 91.2 & 53.4 & 37.8 & 37.0 & 60.7 \\
PNA \cite{wei2022towards} & 61.0 & 56.4 & 53.5 & 95.6 & 91.8 & 52.4 & 38.0 & 34.6 & 60.4 \\
PNA + PathOut \cite{wei2022towards} & 59.6 & 54.2 & 52.4 & 94.2 & 90.2 & 51.8 & 38.4 & 35.8 & 59.5 \\
\citet{cuiRobustnessLargeMultimodal2024} & 60.6 & 53.2 & 54.6 & 97.8 & 90.8 & 54.8 & 37.6 & 36.8 & 60.7 \\
\citet{schlarmannAdversarialRobustnessMultiModal2023} & 58.8 & 52.8 & 54.2 & 93.0 & 90.8 & 51.0 & 38.0 & 34.8 & 62.6 \\
Attack-Bard~\cite{dongHowRobustGoogles2023} & 63.2 & 55.0 & 56.0 & 97.4 & 92.4 & 55.8 & 41.4 & 38.0 & 62.4 \\
VT-Attack~\cite{wangBreakVisualPerception2024} & 61.6 & 55.6 & 56.6 & 98.2 & 91.0 & 54.2 & 39.8 & 37.6 & 61.7 \\
\textbf{\AttackName} & 64.2 & 58.4 & 60.2 & 98.2 & 93.6 & 56.2 & 43.0 & 38.0 & \textbf{64.5} \\
\bottomrule
\end{tabular}}
\end{table*}

\begin{table*}[t!]
	\caption{Comparison of attack imperceptibility.}
    \label{Fig:stealthiness}
	\centering

	\resizebox{\linewidth}{!}{
		\begin{tabular}{l|cccccc}
			\toprule
			\textbf{Attack} & \textbf{SSIM} $\uparrow$ & \textbf{LPIPS} $\downarrow$ & \textbf{MS-SSIM} $\uparrow$& \textbf{FSIM} $\uparrow$& \textbf{VIF} $\uparrow$ & \textbf{HaarPSI} $\uparrow$ \\
			\midrule
            Clean &  1 & 0 & 1 & 1 & 1 & 1  \\
            TGR \cite{zhangTransferable2023}&0.8943&0.0428&0.9828&0.7967&0.6383&0.9387\\
PNA \cite{wei2022towards}&0.8596& 0.0619&0.9743&0.7835&0.5933&0.9132\\
PNA + PathOut \cite{wei2022towards}&0.9086&0.0379&0.9832&\textbf{0.8320}&0.6852&0.9401 \\
\citet{cuiRobustnessLargeMultimodal2024}&0.8978&0.0406&0.9835& 0.7991&0.6444&0.9395 \\
            \citet{schlarmannAdversarialRobustnessMultiModal2023} &0.9063&0.0378&0.9856&0.8079&0.6604&\textbf{0.9450}\\
            Attack-Bard~\cite{dongHowRobustGoogles2023}& 0.8924&0.0397&0.9833&0.7976&0.6433&0.9393 \\
		VT-Attack \cite{wangBreakVisualPerception2024} & 0.8978&0.0411&0.9836&0.7988&0.6440&0.9392 \\
            \rowcolor{gray!15}
             \textbf{\AttackName}  &\textbf{0.9161}&\textbf{0.0369}&\textbf{0.9857}&0.8095&\textbf{0.6775}&0.9336\\
			\bottomrule
	\end{tabular}}
\end{table*}

\section{Detailed Experimental Setup}
\label{sec:additional_setup}

\noindent\textbf{Evaluation Attacks.}
We evaluate four representative untargeted attacks on LVLMs, including one end-to-end and three encoder-based approaches:

\begin{itemize}[leftmargin=*, itemsep=2pt, topsep=0pt, parsep=0pt]
    \item \citet{schlarmannAdversarialRobustnessMultiModal2023} introduce an end-to-end white-box attack that generates adversarial examples by minimizing the cross-entropy loss between the original and adversarial outputs of the entire LVLM. We adapt it to a black-box setting, where adversarial examples are crafted on BLIP-2 and then transferred to other LVLMs.
    
    \item \citet{cuiRobustnessLargeMultimodal2024} design an encoder-based attack that generates adversarial examples by minimizing the cosine similarity between visual features of the adversarial image and the corresponding text embedding.
    
    \item Attack-Bard~\cite{dongHowRobustGoogles2023} is an encoder-based attack that maximizes the distance between clean and adversarial visual features, pushing the adversarial image representation away from the original visual backbone encoding.
    
    \item VT-Attack~\cite{wangBreakVisualPerception2024} is an encoder-based attack that perturbs visual tokens from multiple perspectives, including local feature representations, inter-token relationships, and global semantics. This multi-faceted disruption is designed to more comprehensively break cross-modal alignment in LVLMs.
\end{itemize}

\noindent\textbf{LVLM-As-a-Judge.}
We use the following prompt to determine correctness.
\begin{promptbox}
You are given a description: \textit{\#Description} \\
Carefully observe the provided image. Your task is to answer the following question clearly and precisely:

Can the description be reasonably used to describe the content of the image, even if it does not cover all objects or details? Answer "Yes" if the description is a plausible and relevant description of the image as a whole. Otherwise, answer "No".

Answer in the following format:
\texttt{Match with image: <Yes/No>}
\end{promptbox}
When the model outputs ``\texttt{Match with image: No}'', it indicates that the adversarial description no longer aligns with the image content. We then compute the attack success rate (ASR) as the fraction of adversarial examples judged as \texttt{No} over the total number of evaluated cases. 

\begin{remark}
We adopt an LVLM-as-a-judge protocol because image captioning is an open-ended, one-to-many task: a single image admits many valid descriptions, and lexical-overlap metrics can be poorly aligned with whether a caption is \emph{faithful} to the image.
An LVLM judge can directly assess semantic and visual consistency, and prior work has shown that GPT-4V can serve as a general-purpose evaluator~\cite{zhang2023gpt}.
Our evaluation is designed to mitigate potential judge-induced biases in two ways:
(1) The judge is provided with the clean image and the adversarial/clean model outputs only, and never observes the adversarial image, avoiding being confounded by the perturbation itself;
(2) We use the judge LVLM that is architecturally different from the victim model to reduce same-model bias. Recent work also confirms that such LVLM-based evaluation protocols are reliable and demonstrate high alignment with human judgment~\cite{10.5555/3692070.3692324}. We also provide a detailed analysis of the evaluation protocol in \Cref{sec:additional_ablaction_study}.
\end{remark}

\noindent\textbf{Surrogate and Victim Models.} For the surrogate model, we adopt the widely used CLIP ViT-L/14 for all encoder-based attacks~\cite{zhaoEvaluatingAdversarialRobustness2023, wangBreakVisualPerception2024}, while BLIP-2 serves as the surrogate model for end-to-end attacks. For victim models, we select several representative LVLMs with diverse visual encoders. Specifically, both LLaVA \cite{liuVisualInstructionTuning2023} and OpenFlamingo \cite{alayracFlamingoVisualLanguage2022} adopt CLIP ViT-L/14 \cite{radfordLearningTransferableVisual2021a}, while BLIP-2 \cite{liBLIP2BootstrappingLanguageImage2023} uses EVA-CLIP \cite{fangEVAExploringLimits2023}. Qwen2.5-VL \cite{qwen2.5-VL} uses a native dynamic resolution ViT tailored for high-resolution understanding. InternVL3 \cite{zhu2025internvl3exploringadvancedtraining} is built upon InternViT, a hierarchical vision backbone optimized for multimodal fusion. Kimi-VL \cite{kimiteam2025kimivltechnicalreport} leverages MoonViT, which supports native-resolution input and efficient scaling via a mixture-of-experts framework. Additionally, we include several popular commercial black-box LVLMs, GPT-4o~\cite{openai2024gpt4technicalreport}, GPT-5.4, and Gemini 2.0 Flash~\cite{team2023gemini}, to evaluate the real-world applicability in closed-source scenarios.
By including models with both similar (e.g., ViT-L/14 variants) and distinct (e.g., MoonViT, interViT) visual encoders and different LLMs, we aim to comprehensively assess transferability
\Cref{tab:victim_model} gives the victim models and the corresponding link for the reproduction of the results. 

\noindent\textbf{Prompt Details.}  
For image classification on CIFAR-10, we use the following instruction prompt:  

\begin{promptbox}
You are an image classifier. \\
Given an image, classify it into exactly one of these 10 categories: airplane | automobile | bird | cat | deer | dog | frog | horse | ship | truck. \\
Respond with only the category name.
\end{promptbox}

For OpenFlamingo, we instead adopt the shorter template ``a photo of a'', as its instruction-following capability is relatively limited.

\section{Additional Experimental Results}
\label{sec:additional_exmperimental_results}
\subsection{Additional Attack Performance}
\label{app:full_results}

To provide a more comprehensive evaluation, we include additional results on diverse LVLMs and tasks beyond the main paper.

\noindent\textbf{Attack Performance on More LVLMs.} \Cref{tab:more_lvllm_results} reports the attack performance on several open-source LVLMs not covered in the main tables. These include models with more diverse architectures, parameter scales. The results further verify that \AttackName maintains strong transferability and semantic disruption capabilities across a broad range of black-box settings.

\noindent\textbf{Attack Performance on Other Tasks.}
\Cref{tab:CLS_VQA} presents the performance of the proposed attack under different multimodal tasks, including visual question answering (VQA) and captioning. We observe that while captioning tasks often rely on patch-level information, VQA performance is more sensitive to subtle region–text alignments, which are effectively disrupted by our method. These results confirm that \AttackName is versatile and effective across various task types.

\subsection{Attack Imperceptibility} 
\label{sec:attack_imperceptiblity}
To assess human imperceptibility, we adopt widely used full-reference image quality metrics.
Higher values of SSIM~\cite{wang2004ssim}, MS-SSIM~\cite{wang2003multiscale}, FSIM~\cite{zhang2011fsim}, VIF~\cite{sheikh2006vif}, and HaarPSI~\cite{reisenhofer2018haarpsi} indicate better preservation of structural, feature, and visual fidelity relative to the clean image, while lower LPIPS~\cite{zhang2018lpips} values indicate smaller perceptual differences in deep feature space.
As shown in \Cref{Fig:stealthiness}, \AttackName achieves the best performance across most perceptual metrics. The results demonstrate that, despite significantly improving attack effectiveness, our method does not compromise stealthiness. \AttackName explicitly constrains perturbations to semantically important regions. While this introduces slightly more localized changes compared to uniformly distributed or less targeted perturbations, it maintains comparable perceptual quality from a human perspective.

\begin{table}[t]
  \centering
  \small
  \caption{Pairwise Q1 agreement (``match with image'') on 100 sampled examples. \emph{Gemini-2.5-fl.-img.}\ denotes \texttt{gemini-2.5-flash-image}; \emph{Gemini-3-fl.-prev.}\ denotes \texttt{gemini-3-flash-preview}.}
  \label{tab:judge_q1_pairwise}
  \begin{tabular}{@{}llc@{}}
    \toprule
    Judge A & Judge B & Agree (\%) \\
    \midrule
    GPT-4.1 & GPT-4o                 & 98 \\
    GPT-4.1 & Gemini-2.5-fl.-img.    & 90 \\
    GPT-4.1 & Gemini-3-fl.-prev.     & 90 \\
    GPT-4o  & Gemini-2.5-fl.-img.    & 92 \\
    GPT-4o  & Gemini-3-fl.-prev.     & 88 \\
    Gemini-2.5-fl.-img. & Gemini-3-fl.-prev. & 84 \\
    \midrule
    \multicolumn{2}{@{}l}{Mean over 6 pairs} & \textbf{91} \\
    \bottomrule
  \end{tabular}
\end{table}

\begin{table*}[t!]
\caption{
Effect of different surrogate models. \textbf{Bold} denotes best performance. 
}
\label{tab:surrogate_effect}
\centering
\resizebox{\linewidth}{!}{
\begin{tabular}{ll|cccccc|c}
\toprule
\multirow{2}{*}{\textbf{Surrogate}} & \multirow{2}{*}{\textbf{Attack}} & \multicolumn{6}{c|}{\textbf{CLIP Similarity between image and generated text $\downarrow$}} & \multirow{2}{*}{\textbf{ASR (\%) $\uparrow$}}\\
&& RN-50 & RN-101 & ViT-B/16 & ViT-B/32 & ViT-L/14 & Ensemble & \\
\midrule
\multirow{4}{*}{CLIP-L/14}
&\citet{cuiRobustnessLargeMultimodal2024}&0.2365 & 0.4584 & 0.2981 & 0.2925 & 0.2530 & 0.3077& 41.8\\
&Attack-Bard~\cite{dongHowRobustGoogles2023}&0.2354 & 0.4568 & 0.2964 & 0.2915 & 0.2498 & 0.3060&38.4\\
&VT-Attack \cite{wangBreakVisualPerception2024} &0.2330 & 0.4544 & 0.2939 & 0.2892 & 0.2462 & 0.3033&46.0  \\
& \textbf{\AttackName}  &{0.2282} & {0.4493} & {0.2873} & {0.2831} & {0.2376} & {0.2971} &\textbf{55.4}  \\ 
\midrule
\multirow{4}{*}{CLIP-B/16}
& \citet{cuiRobustnessLargeMultimodal2024} & 0.2410 & 0.4624 & 0.3042 & 0.2991 & 0.2592 & 0.3132 & 28.4\\
& Attack-Bard~\cite{dongHowRobustGoogles2023} & 0.2424 & 0.4644 & 0.3040 & 0.2998 & 0.2610 & 0.3143& 21.0\\
& VT-Attack~\cite{wangBreakVisualPerception2024} & 0.2410 & 0.4624 & 0.3042 & 0.2991 & 0.2592 & 0.3132 & 28.4 \\
& \textbf{\AttackName} &0.2396 & 0.4599 & 0.2999 & 0.2964 & 0.2560 & 0.3104& \textbf{36.6}\\
\midrule
\multirow{4}{*}{SigLIP}
& \citet{cuiRobustnessLargeMultimodal2024} & 0.2416 & 0.4642 & 0.3058 & 0.3003 & 0.2616 & 0.3147 & 19.8\\
& Attack-Bard~\cite{dongHowRobustGoogles2023} &0.2409 & 0.4625 & 0.3039 & 0.2991 & 0.2606 & 0.3134& 24.6\\
& VT-Attack~\cite{wangBreakVisualPerception2024} &0.2407 & 0.4614 & 0.3029 & 0.2978 & 0.2579 & 0.3121 & 28.4\\
& \textbf{\AttackName} & 0.2388 & 0.4614 & 0.3028 & 0.2969 & 0.2581 & 0.3116 & \textbf{34.2}\\
\midrule
\multirow{4}{*}{DINOv2-B}
& \citet{cuiRobustnessLargeMultimodal2024} & 0.2438 & 0.4666 & 0.3083 & 0.3023 & 0.2667 & 0.3175 & 16.0\\
& Attack-Bard~\cite{dongHowRobustGoogles2023} & {0.2429} & {0.4653} & {0.3068} & {0.3018} & {0.2650} & {0.3164} & 16.4\\
& VT-Attack~\cite{wangBreakVisualPerception2024} & 0.2436 & 0.4661 & 0.3077 & 0.3015 & 0.2659 & 0.3169 & 19.2\\
& \textbf{\AttackName} & 0.2428 & 0.4650 & 0.3070 & 0.3012 & 0.2643 & 0.3161 & \textbf{19.6}\\
\bottomrule
\end{tabular}}
\end{table*}

\begin{table*}[t!]
\caption{Effect of base ratio $r$.}
\label{fig:Budget_Ration}
\centering

\resizebox{\linewidth}{!}{
\begin{tabular}{c|c|cccccc|c}
\toprule
\multirow{2}{*}{$r$}&\multirow{2}{*}{\textbf{Victim LVLM}}
& \multicolumn{6}{c|}{\textbf{CLIP Similarity Score between image and generated text ($\downarrow$)}} & \multirow{2}{*}{\textbf{ASR (\%) ($\uparrow$)}}\\
& &  RN-50 & RN-101 & ViT-B/16 & ViT-B/32 & ViT-L/14 & Ensemble  &  \\
\midrule
\multirow{3}{*}{\textbf{0.2}}
&LLaVA& 0.2279 & 0.4481 & 0.2861 & 0.2822 & 0.2347 & \textbf{0.2958} & 56.0\\
&Qwen2.5-VL& 0.2468 & 0.4724 & 0.3064 & 0.3010 & 0.2524 & 0.3158 & 37.8\\
&InternVL3& 0.2460 & 0.4720 & 0.3038 & 0.3010 & 0.2551 & \textbf{0.3156} & 41.4\\
\midrule
\multirow{3}{*}{\textbf{0.4}}
&LLaVA& 0.2302 & 0.4495 & 0.2880 & 0.2846 & 0.2369 & 0.2978 & 55.4\\
&Qwen2.5-VL& 0.2468 & 0.4732 & 0.3052 & 0.3014 & 0.2516 & \textbf{0.3156} & 39.0\\
&InternVL3& 0.2462 & 0.4718 & 0.3052 & 0.3013 & 0.2559 & 0.3161 & 41.2\\
\midrule
\multirow{3}{*}{\textbf{0.6}}
&LLaVA& 0.2302 & 0.4511 & 0.2891 & 0.2848 & 0.2383 & 0.2987 & 51.2\\
&Qwen2.5-VL& 0.2483 & 0.4743 & 0.3077 & 0.3037 & 0.2551 & 0.3178 & 36.4\\
&InternVL3& 0.2498 & 0.4741 & 0.3075 & 0.3046 & 0.2586 & 0.3190 & 38.6\\
\midrule
\multirow{3}{*}{\textbf{0.8}}
&LLaVA& 0.2305 & 0.4516 & 0.2893 & 0.2861 & 0.2404 & 0.2996 & 50.4\\
&Qwen2.5-VL&0.2490 & 0.4758 & 0.3093 & 0.3046 & 0.2562 & 0.3190 & 34.6\\
&InternVL3& 0.2499 & 0.4751 & 0.3083 & 0.3030 & 0.2592 & 0.3191 & 39.0\\
\midrule
\multirow{3}{*}{\textbf{1.0}}
&LLaVA& 0.2313 & 0.4527 & 0.2923 & 0.2872 & 0.2442 & 0.3015 & 48.8\\
&Qwen2.5-VL&0.2512 & 0.4772 & 0.3117 & 0.3062 & 0.2586 & 0.3209 & 31.2\\
&InternVL3& 0.2511 & 0.4770 & 0.3102 & 0.3059 & 0.2621 & 0.3213 & 35.4\\
\bottomrule
\end{tabular}}
\end{table*}




\begin{table*}[t!]
\caption{Effect of relevance threshold $\tau$.}
\label{fig:relevance_threshold}
\centering

\resizebox{\linewidth}{!}{
\begin{tabular}{c|c|cccccc|c}
\toprule
\multirow{2}{*}{$\tau$}&\multirow{2}{*}{\textbf{Victim LVLM}}
& \multicolumn{6}{c|}{\textbf{CLIP Similarity Score between image and generated text ($\downarrow$)}} & \multirow{2}{*}{\textbf{ASR (\%) ($\uparrow$)}}\\
& &  RN-50 & RN-101 & ViT-B/16 & ViT-B/32 & ViT-L/14 & Ensemble &  \\
\midrule

\multirow{4}{*}{\textbf{0.1}}
&LLaVA& 0.2304 & 0.4513 & 0.2903 & 0.2859 & 0.2405 & 0.2997 & 52.2\\
&Qwen2.5-VL& 0.2483 & 0.4742 & 0.3075 & 0.3027 & 0.2528 & 0.3171& 35.4\\
&InternVL3& 0.2467 & 0.4731 & 0.3064 & 0.3022 & 0.2549 & 0.3166 & 42.8\\
\midrule
\multirow{4}{*}{\textbf{0.2}}
&LLaVA& 0.2293 & 0.4506 & 0.2891 & 0.2850 & 0.2381 & 0.2984 & 54.2\\
&Qwen2.5-VL&  0.2471 & 0.4737 & 0.3064 & 0.3021 & 0.2541 & 0.3167 & 37.0\\
&InternVL3& 0.2475 & 0.4736 & 0.3063 & 0.3020 & 0.2563 & 0.3171 & 38.8\\
\midrule
\multirow{4}{*}{\textbf{0.3}}
&LLaVA& 0.2302 & 0.4495 & 0.2880 & 0.2846 & 0.2369 & 0.2978 & \textbf{55.4}\\
&Qwen2.5-VL& 0.2468 & 0.4732 & 0.3052 & 0.3014 & 0.2516 &\textbf{0.3156} & \textbf{39.0}\\
&InternVL3& 0.2462 & 0.4718 & 0.3052 & 0.3013 & 0.2559 & \textbf{0.3161} & \textbf{41.2}\\
\midrule
\multirow{4}{*}{\textbf{0.4}}
&LLaVA& 0.2292 & 0.4502 & 0.2874 & 0.2844 & 0.2370 & \textbf{0.2977} & 53.8\\
&Qwen2.5-VL& 0.2469 & 0.4737 & 0.3059 & 0.3012 & 0.2524 & 0.3160 & 37.2\\
&InternVL3& 0.2474 & 0.4721 & 0.3057 & 0.3009 & 0.2552 & 0.3163 & 40.8\\
\bottomrule
\end{tabular}}
\end{table*}

\begin{figure}[t]
    \centering
    \includegraphics[width=0.6\linewidth]{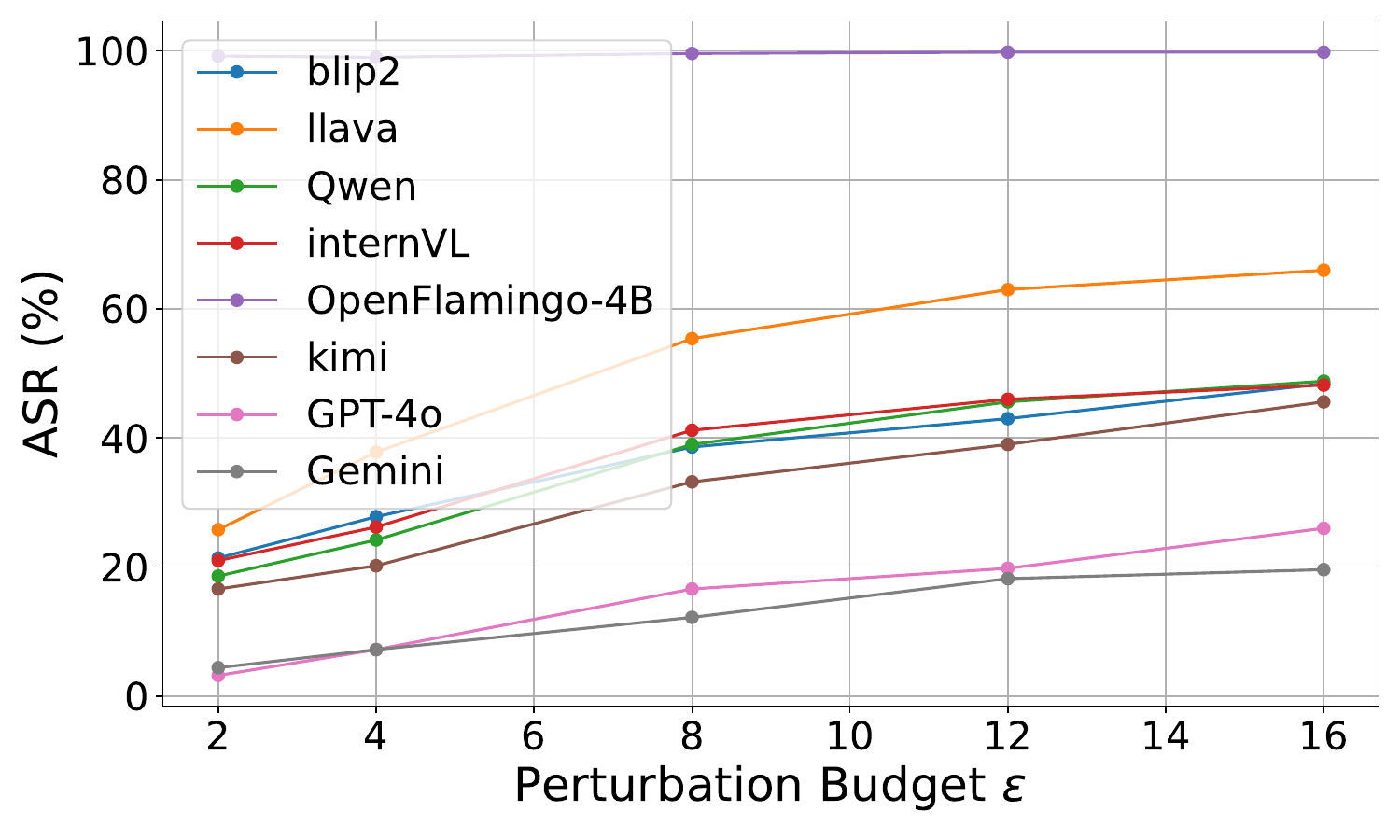}
    \caption{Attack performance across perturbation budgets.}
    \label{fig:perturbation}
\end{figure}

\begin{figure}[t]
    \centering
    \includegraphics[width=0.6\linewidth]{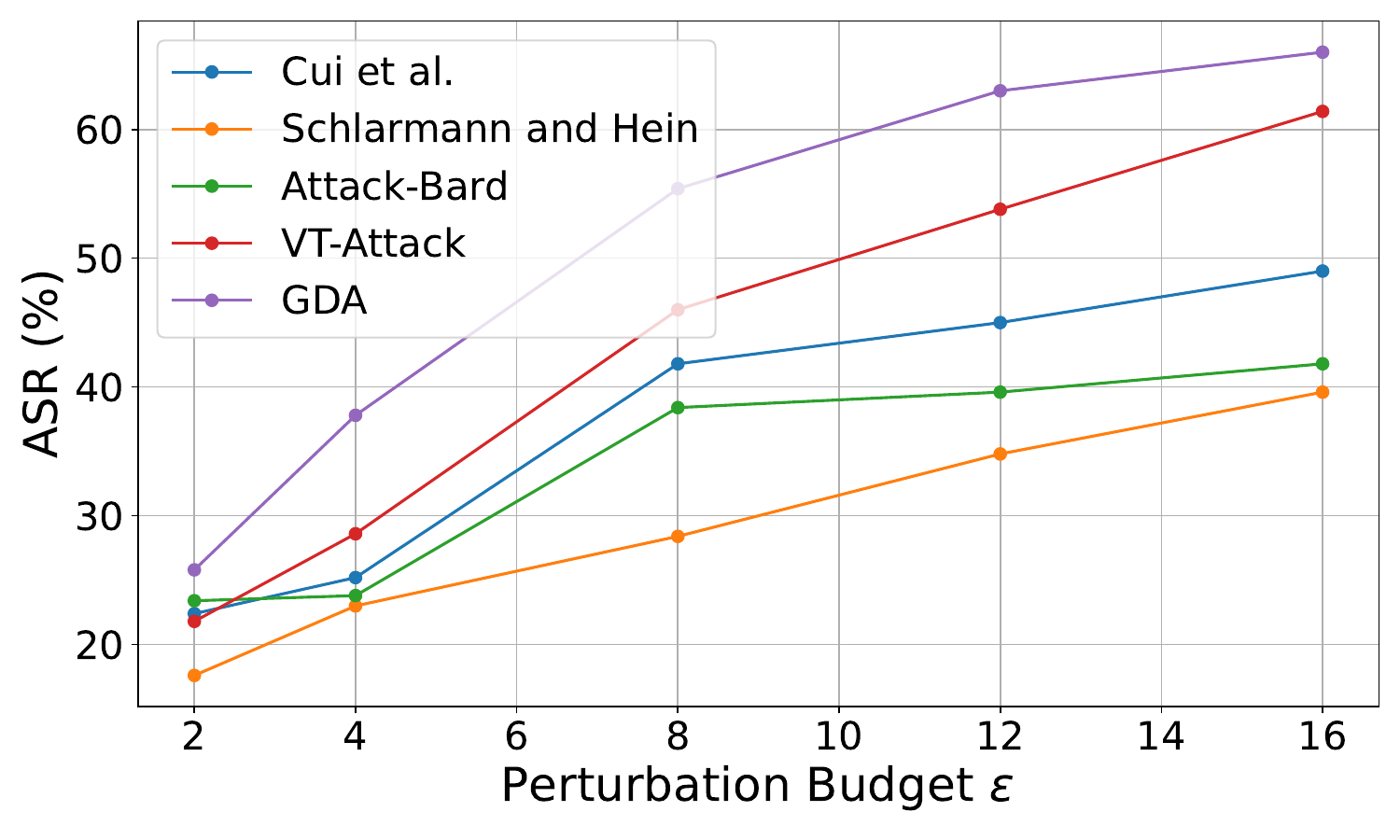}
    \caption{
    ASR (\%) on LLaVA under varying $\ell_\infty$ perturbation budgets $\epsilon \in \{2,4,8,12,16\}$. Our method (\AttackName) consistently outperforms existing encoder-based attacks, with sharper gains at larger $\epsilon$.}
    \label{fig:budget_comparison}
\end{figure}

\begin{figure}[t]
    \centering
    \includegraphics[width=0.6\linewidth]{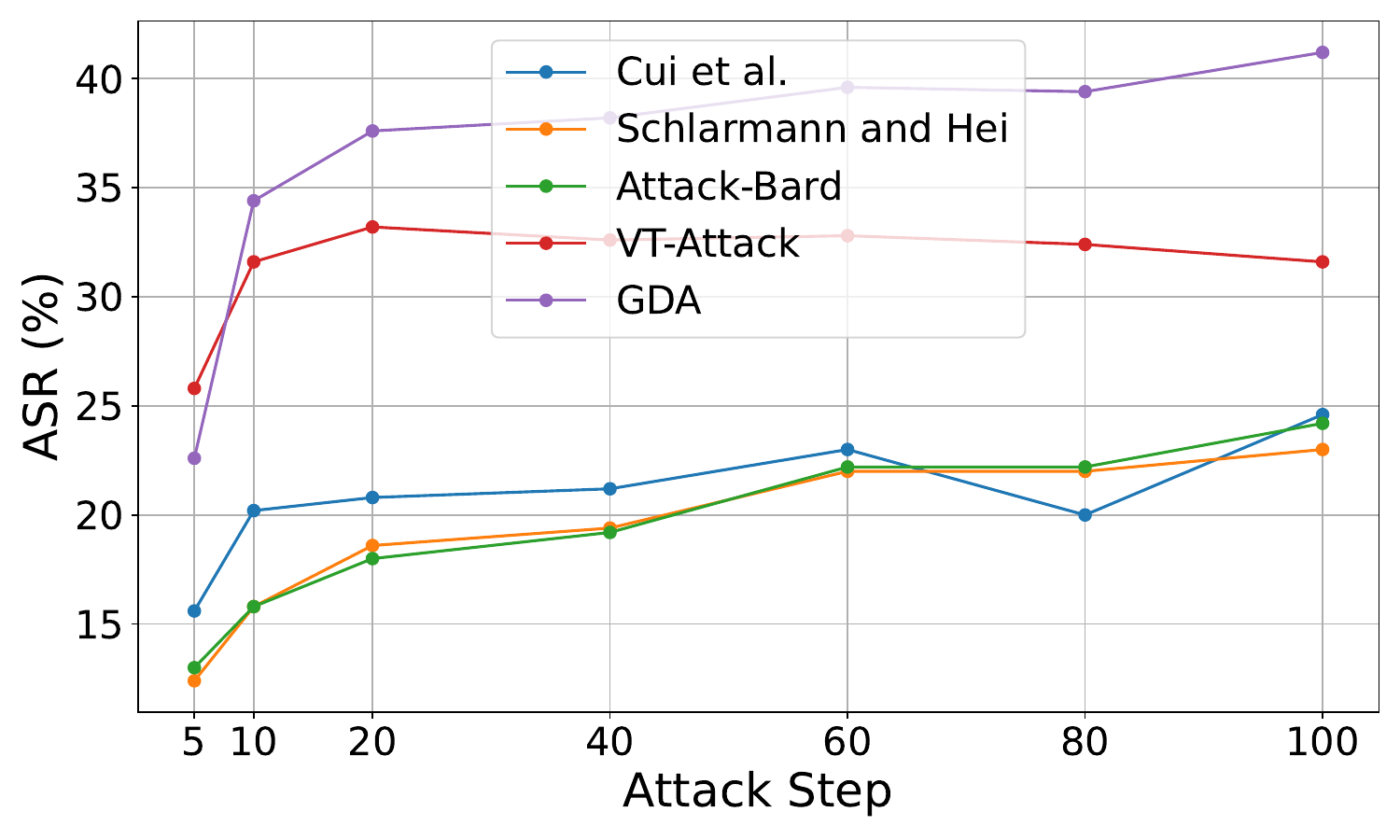}
\caption{
ASR (\%) on InternVL under varying attack steps.
Increasing the number of steps improves ASR for all methods, but \AttackName exhibits more consistent gains and reaches the highest ASR at 100 steps.
}
\label{fig:step_vs_asr}
\end{figure}

\subsection{Additional Ablation Studies and Analysis}
\label{sec:additional_ablaction_study}

\begin{table*}[t]
\centering
\caption{Component ablation of \AttackName. $\checkmark_{\text{II}}$ = global disruption with only image--image loss. Without GPA, all pixels are subject to the same perturbation bound.}
\label{fig:Ablation_stydy_SGMA}
\resizebox{\textwidth}{!}{
\begin{tabular}{c|ccc|cccccc|c}
\toprule
\multirow{2}{*}{\textbf{Victim LVLM}}&
\multicolumn{3}{c|}{\textbf{Method}}
& \multicolumn{6}{c|}{\textbf{CLIP Similarity between image and generated text ($\downarrow$)}} & \multirow{2}{*}{\textbf{ASR (\%) ($\uparrow$)}}\\
&Global disrup. & Local disrup. & GPA
 & RN-50 & RN-101 & ViT-B/16 & ViT-B/32 & ViT-L/14 & Ensemble & \\
\midrule
\multirow{4}{*}{\textbf{LLaVA}}
&$\checkmark_{\text{II}}$&&& 0.2377 & 0.4590 & 0.2988 & 0.2945 & 0.2530 & 0.3086& 39.0 \\
&\checkmark&&&0.2324 & 0.4547 & 0.2935 & 0.2885 & 0.2459 & 0.3030 & 44.2\\
&\checkmark&\checkmark&& 0.2354 & 0.4564 & 0.2971 & 0.2909 & 0.2502 & 0.3060 & 48.8\\
&\checkmark&\checkmark&\checkmark&0.2302 & 0.4495 & 0.2880 & 0.2846 & 0.2369 & 0.2978 & 55.4\\
\midrule
\multirow{4}{*}{\textbf{Qwen2.5-VL}}
&$\checkmark_{\text{II}}$&&& 0.2536 & 0.4786 & 0.3143 & 0.3090 & 0.2620 & 0.3235& 25.0 \\
&\checkmark&&&0.2528 & 0.4766 & 0.3117 & 0.3063 & 0.2603 & 0.3215 & 30.0\\
&\checkmark&\checkmark&& 0.2512 & 0.4772 & 0.3117 & 0.3062 & 0.2586 & 0.3209 & 31.2\\
&\checkmark&\checkmark&\checkmark&0.2468 & 0.4732 & 0.3052 & 0.3014 & 0.2516 & 0.3156 & 39.0\\
\midrule
\multirow{4}{*}{\textbf{InternVL3}}
&$\checkmark_{\text{II}}$&&& 0.2553 & 0.4804 & 0.3156 & 0.3108 & 0.2676 & 0.3260& 23.5 \\
&\checkmark&&& 0.2507 & 0.4774 & 0.3119 & 0.3055 & 0.2632 & 0.3217 & 28.7\\
&\checkmark&\checkmark&&0.2511 & 0.4770 & 0.3102 & 0.3059 & 0.2621 & 0.3213 & 35.4\\
&\checkmark&\checkmark&\checkmark& 0.2462 & 0.4718 & 0.3052 & 0.3013 & 0.2559 & 0.3161 & 41.2\\
\bottomrule
\end{tabular}}
\end{table*}

\noindent\textbf{Component Ablation of \AttackName.}
\Cref{fig:Ablation_stydy_SGMA} presents an ablation analysis of \AttackName on three LVLMs. We progressively introduce global disruption, local disruption, and GPA. The results show that each component contributes to improving attack effectiveness. In particular, adding local disruption on top of global disruption yields a consistent ASR gain (e.g., 44.2\% $\rightarrow$ 48.8\% on LLaVA), while introducing GPA further boosts performance substantially (up to 55.4\% on LLaVA).

\begin{table}[t]
\centering
\caption{Comparison of runtime efficiency.
We report the average runtime (seconds per sample) required to generate an adversarial example. 
}
\label{tab:runtime}
\begin{tabular}{lc}
\toprule
\textbf{Attack Method} & \textbf{Time (s/sample)} \\
\midrule
\citet{schlarmannAdversarialRobustnessMultiModal2023} & 28.4024 \\
\midrule
\citet{cuiRobustnessLargeMultimodal2024}      & \phantom{0}5.5649 \\
Attack-Bard~\cite{dongHowRobustGoogles2023}    & \phantom{0}5.8810 \\
VT-Attack~\cite{wangBreakVisualPerception2024}        & \phantom{0}9.7486 \\
\midrule
\textbf{\AttackName}          & 10.3353 \\
\bottomrule
\end{tabular}%
\end{table}

\noindent\textbf{Robustness of LVLM-as-a-Judge.}
Using a single proprietary LVLM as an automatic judge may raise concerns about \emph{judge-specific bias}: the reported attack success or semantic judgments may partly reflect the idiosyncrasies of one API rather than a stable evaluation protocol.
To address this concern, we quantify \textbf{cross-model consistency} under a fixed prompting scheme, using the same image, the same clean and adversarial captions, identical instructions, and structured Yes/No outputs.
We report pairwise agreement across different judges.
These metrics measure \textbf{consistency among automated judges}, rather than accuracy with respect to human annotations, which we do not claim at scale.

We first evaluate whether the adversarial caption is still judged to be a plausible description of the image (Q1).
Table~\ref{tab:judge_q1_pairwise} summarizes pairwise agreement on 100 randomly sampled attack examples (seed $7$), using clean captions and adversarial LLaVA captions under the same judging prompt.
GPT-4.1 and GPT-4o are highly consistent on Q1, reaching 98\% agreement.
Both Gemini-family judges also show strong agreement with GPT-4.1, at around $90\%$ pairwise agreement.
Agreement between the two Gemini variants is somewhat lower (84\%), which is expected when comparing judges across vendors and model families.
Across all six unordered pairs of the four judges, the mean pairwise Q1 agreement reaches \textbf{90.2\%}.
Moreover, for 82 out of the 100 sampled examples, all judges with valid Q1 labels produce the same decision, yielding a unanimous-agreement rate of $82\%$.

\noindent\textbf{Effect of Surrogate Models.}
We further investigate how the choice of surrogate vision encoder affects the transferability of encoder-based attacks. 
Since LLaVA uses CLIP-L/14 as its visual backbone, attacks optimized on the same CLIP-L/14 surrogate achieve the strongest transferability across all methods. 
As shown in \Cref{tab:surrogate_effect}, replacing the surrogate with CLIP-B/16, SigLIP, or DINOv2-B consistently reduces the ASR, indicating that architectural and representation mismatch between the surrogate and victim substantially weakens black-box transferability.
Despite this degradation, \AttackName consistently achieves the best performance under all surrogate settings. 
With the matched CLIP-L/14 surrogate, \AttackName improves the ASR from 46.0\% to 55.4\% over VT-Attack, demonstrating the effectiveness of grounding-aware perturbation allocation when the surrogate provides well-aligned visual representations. 
When using weaker or mismatched surrogates, the advantage of \AttackName remains observable: it improves the ASR from 28.4\% to 36.6\% under CLIP-B/16 and from 28.4\% to 34.2\% under SigLIP. 
Even under the highly mismatched DINOv2-B surrogate, where all attacks suffer from limited transferability, \AttackName still achieves the highest ASR of 19.6\%, slightly outperforming VT-Attack.
These results suggest two findings. 
First, surrogate-victim alignment is a critical factor for encoder-based adversarial transfer, as attacks optimized on representations closer to the victim backbone are more effective. 
Second, the gain of \AttackName is not solely tied to CLIP-family surrogates. 
Although strong architectural mismatch inevitably limits the overall attack strength, explicitly allocating perturbations to text-grounded visual evidence still provides a consistent improvement over prior encoder-based attacks.

\noindent\textbf{Effect of Perturbation Budget $\epsilon$.} We evaluate the performance of \AttackName under varying perturbation budgets $\epsilon \in \{2, 4, 8, 12, 16\}$. As illustrated in \Cref{fig:perturbation}, the ASR consistently increases with the perturbation magnitude across all tested LVLMs. Notably, even under a low budget (e.g., $\epsilon = 4$), \AttackName achieves non-trivial ASR, demonstrating its effectiveness under tight constraints. When $\epsilon = 16$, the ASRs on GPT-4o and Gemini reach up to 20\%, indicating that the attack remains effective even on more robust models. These results underscore the scalability and generalization ability of \AttackName under different perturbation levels. We additionally compare the attack performance under different perturbation budgets on LLaVA in \Cref{fig:budget_comparison}. Across all methods, ASR increases monotonically with $\epsilon$, indicating better attack effectiveness under larger perturbations. 
Notably, \AttackName consistently achieves the highest ASR across all budgets. While most baselines (e.g., Cui et al., Attack-Bard) plateau around 40--50\% when $\epsilon = 16$, \AttackName reaches 66\%, showing its stronger optimization and better alignment with downstream model behavior. These results further demonstrate \AttackName's superior performance.

\noindent\textbf{Computational Cost.}
We evaluate the computational cost of \AttackName by measuring the average runtime per sample under identical hardware and software environments. As shown in \Cref{tab:runtime}, \AttackName demonstrates a significant efficiency advantage over the end-to-end baseline, achieving a $2.7\times$ speedup compared to Schlarmann \& Hein~\cite{schlarmannAdversarialRobustnessMultiModal2023} (10.34s vs. 28.40s). This efficiency stems from our strategy of targeting the vision encoder rather than optimizing the entire VLM pipeline. Compared to other transfer-based attacks (e.g., Cui et al.~\cite{cuiRobustnessLargeMultimodal2024} and Attack-Bard~\cite{dongHowRobustGoogles2023}), \AttackName incurs a marginal increase in computational overhead. This is primarily due to the calculation of the semantic relevance location. Crucially, this additional computation is a one-time cost per image, independent of the number of optimization steps, ensuring scalability. While \AttackName is comparable to VT-Attack~\cite{wangBreakVisualPerception2024} in runtime, it offers superior transferability. 

\noindent\textbf{Generalization to Other Tasks.}
In order to evaluate \AttackName more comprehensively, we introduce two additional tasks: image classification and VQA. As shown in \Cref{tab:CLS_VQA} in Appendix \ref{app:full_results}, our attack achieves consistently higher ASR across all LVLMs compared to prior methods. For image classification, \AttackName attains the highest average ASR of 60.7\%, surpassing all baselines. On the VQA task, \AttackName achieves the top average ASR of 61.7\%, demonstrating strong robustness across diverse reasoning paradigms. Besides, our perturbations are task-agnostic, generated from the full descriptive caption $T_d$ rather than tailored to specific tasks. While task-specific perturbations that focus on task-relevant regions could further increase ASR, our task-agnostic design demonstrates consistent improvements across different tasks, indicating that \AttackName effectively disrupts fundamental vision–language grounding.

\noindent\textbf{Effect of Base Ratio $r$.}
We analyze the impact of the base perturbation ratio $r$ in the pixel-wise perturbation allocation mechanism (\Cref{eq:perturbation}). As shown in \Cref{fig:Budget_Ration}, a smaller $r$ allocates more perturbation budget to semantically important regions, resulting in lower CLIP scores and higher ASRs. For example, when $r$ decreases from 1.0 to 0.2, the ASR on Qwen2.5-VL improves from 31.2\% to 37.8\%, while the ensemble CLIP score drops from 0.3209 to 0.3158. However, excessively small $r$ may lead to underutilization of the global image space, limiting robustness. In our experiments, we observe that $r = 0.2$ offers the best trade-off between focused perturbation and overall coverage, and we adopt it as the default setting for \AttackName.

\noindent\textbf{Effect of Relevance Threshold $\tau$.}
We evaluate four values of the relevance threshold $\tau \in \{0.1, 0.2, 0.3, 0.4\}$ to understand its effect on attack performance. A smaller $\tau$ includes more regions, potentially introducing irrelevant areas and diluting perturbation effectiveness. In contrast, a larger $\tau$ selects fewer patches, which may hinder transferability due to overly concentrated perturbations. As shown in \Cref{fig:relevance_threshold}, $\tau=0.3$ consistently achieves the best performance, and is thus adopted as our default setting.

\noindent\textbf{Effect of Attack Steps.} Adhering to the evaluation principles established in~\cite{carlini2019evaluating}, we analyze the evolution of ASR over varying PGD steps to assess convergence. As illustrated in \Cref{fig:step_vs_asr}, ASR improves with the step count for all methods before gradually saturating, indicating stable convergence under a fixed perturbation budget. While baseline methods (e.g., \cite{cuiRobustnessLargeMultimodal2024,schlarmannAdversarialRobustnessMultiModal2023}) plateau relatively early, showing negligible gains beyond 40 steps, \AttackName consistently achieves superior ASR across all intervals. It maintains moderate improvements as optimization proceeds, a trajectory that suggests more effective gradient guidance rather than delayed convergence. To ensure fair comparison and eliminate optimization-related confounds, we standardize on 100 PGD steps for all main experiments, ensuring all methods are evaluated well within their stable performance regimes.

\begin{table}
\centering
\small
\caption{Effect of adding verbs to the local grounding term. Results are reported on LLaVA. Lower CLIP similarity and higher ASR indicate a stronger attack.}
\label{tab:verb_ablation}
\resizebox{\linewidth}{!}{
\begin{tabular}{l|cccccc|c}
\toprule
\textbf{Local Grounding Unit}
& RN-50 & RN-101 & ViT-B/16 & ViT-B/32 & ViT-L/14 & Ensemble & \textbf{ASR (\%) $\uparrow$} \\
\midrule
Noun phrases only & 0.2302 & 0.4495 & 0.2880 & 0.2846 & 0.2369 & 0.2978 & \textbf{55.4} \\
Noun phrases + verbs & 0.2299 & 0.4485 & 0.2870 & 0.2843 & 0.2364 & 0.2972 & 54.3 \\
\bottomrule
\end{tabular}}
\end{table}

\noindent\textbf{Effect of Adding Verbs.} As shown in \Cref{tab:verb_ablation}, extending the local grounding term from noun phrases to noun phrases plus verbs does not improve attack performance. On LLaVA, the ASR drops slightly from 55.4\% to 54.3\%, while the ensemble CLIP similarity changes only marginally from 0.2978 to 0.2972. This suggests that adding verbs does not provide a meaningful benefit in overall attack transferability. A likely reason is that verbs are usually grounded through the involved objects, whereas noun phrases correspond more directly to localized visual entities and are therefore more reliable for patch-level alignment. By contrast, verbs, attributes, and spatial relations are often more distributed and harder to localize precisely at the patch level. This is also why \AttackName uses noun phrases only for the local grounding term, while the global image-text and image-image objectives still perturb overall semantics beyond nouns alone.

\begin{table}[t]
\centering
\small
\caption{Sensitivity of \AttackName to caption quality in Step 1. Results are reported on LLaVA.}
\label{tab:caption_quality}
\resizebox{\linewidth}{!}{
\begin{tabular}{l|cccccc|c}
\toprule
\textbf{Caption Variant}
& RN-50 & RN-101 & ViT-B/16 & ViT-B/32 & ViT-L/14 & Ensemble & \textbf{ASR (\%) $\uparrow$} \\
\midrule
GPT-4o & 0.2302 & 0.4495 & 0.2880 & 0.2846 & 0.2369 & 0.2978 & \textbf{55.4} \\
BLIP-2 caption & 0.2301 & 0.4496 & 0.2881 & 0.2845 & 0.2370 & 0.2979 & 55.3\\
\bottomrule
\end{tabular}}
\end{table}

\noindent\textbf{Effect of Caption Quality.} \Cref{tab:caption_quality} shows that \AttackName is largely insensitive to the specific captioner used in Step 1. Replacing GPT-4o with a smaller open-source captioner such as BLIP-2 changes the ensemble CLIP similarity only marginally, from 0.2978 to 0.2979, and the ASR only slightly, from 55.4\% to 55.3\%. This suggests that the semantic description in \AttackName functions only as a coarse reference for identifying text-grounded evidence, rather than as a target that must be perfectly faithful. In practice, moderate omissions or wording differences are usually tolerable as long as the caption still preserves the main noun phrases corresponding to the key visual entities. In contrast, severe description errors that remove or distort those core entities are more harmful, because they weaken the patch-level grounding signal used by the local objective. This again supports our design choice: \AttackName relies on captions only for coarse grounding, while the global image-text and image-image objectives maintain robustness beyond any single caption realization.

\input{Section/7_Discussion}

%% file: Section/Related_work.tex
\textbf{Adversarial Attacks on LVLMs.} 
With the growing deployment of LVLMs, their vulnerability to adversarial image examples has attracted increasing attention. 
Early studies primarily adopt end-to-end white-box optimization, directly backpropagating through the entire model to manipulate outputs. 
For example, \citet{schlarmannAdversarialRobustnessMultiModal2023} minimize output-level losses to craft perturbations. 
While effective, these methods require full access to the model parameters and are computationally expensive, limiting their practicality. 

To improve efficiency, subsequent work has shifted toward encoder-based attacks, which perturb only the vision encoder. 
Representative approaches include minimizing the cosine similarity between adversarial features and target text embeddings~\cite{cuiRobustnessLargeMultimodal2024}, maximizing the distance between clean and adversarial representations~\cite{dongHowRobustGoogles2023}, and disrupting patch-wise, relational, and global semantics via multi-level objectives~\cite{wangBreakVisualPerception2024}. 
These methods are more lightweight and generalizable, but most of them implicitly assume that the surrogate encoder used in crafting perturbations is identical or similar to the one employed in the victim LVLM. 
Such an assumption is increasingly unrealistic, as modern LVLMs adopt diverse and often proprietary visual backbones and fusion mechanisms. 

To date, only a few studies have attempted to enhance encoder-based transferability under black-box settings. For example, \citet{zhaoEvaluatingAdversarialRobustness2023} refine perturbations using feedback from the victim model, though at the cost of high query overhead and detection risk. 
\citet{xieChainAttackRobustness2025} employ an auxiliary captioning model to produce semantic labels as intermediate targets, which serve as auxiliary signals for crafting perturbations. However, this method still relies on partial architectural or semantic alignment, and effectively assumes that the surrogate vision encoder is similar to that of the victim LVLM. \citet{dongHowRobustGoogles2023} adopt model ensembling, a widely used strategy to boost adversarial transferability across vision models, to improve cross-encoder alignment.
However, despite such efforts, encoder-based transferability under the zero-query black-box setting remains largely unexplored. In this work, we conduct a systematic investigation to address this gap.

\noindent\textbf{Transfer-based Attacks.}
Prior research on transfer-based adversarial attacks in vision tasks has proposed several strategies to enhance transferability. 
Most approaches are general-purpose and apply across different vision models, including \emph{ensemble-model} attacks that optimize over multiple surrogates~\cite{liu2017delving}, \emph{momentum}-based updates that stabilize and amplify gradient directions~\cite{dong2018boosting}, and \emph{input-transformation} strategies (e.g., random resize/crop, input diversity) that improve robustness to model variations~\cite{xie2019improving}. 
In addition, model-specific techniques have been designed for vision transformers, such as perturbing the [CLS] token~\cite{wei2022towards,zhangTransferable2023}. 
While effective in image classification, these ViT-specific strategies are difficult to extend to LVLMs, which rely on distributed patch-level features for cross-modal alignment rather than a single [CLS] representation. 
In contrast, general-purpose transfer strategies remain broadly compatible with LVLMs, but they do not directly address the unique challenges of encoder-based transferability, which remains largely unexplored. Our work complements these lines by studying encoder-based transferability for LVLMs under zero-query constraints and targeting LVLM-specific factors.

%% file: Section/7_Discussion.tex
\section{Discussion}
\label{sec:discussion}

\subsection{Possible Defenses}
The potential social and security risks of our proposed attacks motivate the exploration of effective defense strategies. Following prior work~\cite{wangTransferableMultimodalAttack2024}, we focus on input-level preprocessing methods that require no access to model internals or downstream labels. Specifically, we evaluate four representative techniques: Bit-Red~\cite{xu2017feature}, JPEG compression~\cite{guo2018countering}, neural representation purification (NRP)~\cite{naseer2020self}, and diffusion-based purification (DiffPure)~\cite{nie2022DiffPure}. As shown in \Cref{Tab:defense_asr}, the diffusion-based DiffPure proves to be the most effective, significantly reducing the average ASR from 43.25\% (no defense) to 25.83\%. JPEG compression follows as the second most effective method, lowering the average ASR to 29.39\%, though its lossy nature may impact clean image quality. NRP achieves moderate improvement with an average ASR of 30.80\%, whereas Bit-Red offers only marginal gains (41.95\%), likely due to its limited ability to remove semantically aligned perturbations. Notably, even with strong defenses like DiffPure, attacks remain partially successful (e.g., 54.0\% ASR on OpenFlamingo), underscoring the challenge of completely mitigating such threats.

Beyond preprocessing, adversarial training is another promising direction~\cite{wang2025double,xhonneux2024efficient,shafahi2019adversarial}. Recent studies have attempted to enhance robustness by adversarially training the vision encoder~\cite{wang2025double,gan2020villa}, but such approaches still require large-scale data and significant computational resources, and their gains often fail to transfer fully to LVLMs. Although we do not adopt adversarial training due to resource constraints, we believe it holds promise and encourage future work on more efficient strategies for LVLMs.

\begin{table*}[t!]
\caption{ASR (\%) under different defenses.}
\label{Tab:defense_asr}
\centering
\renewcommand{\arraystretch}{0.9} 
\resizebox{\textwidth}{!}{
\begin{tabular}{l|cccccccc|c}
\toprule
\textbf{Defense}      & \textbf{LLaVA} & \textbf{Qwen2.5-VL} & \textbf{InternVL3} & \textbf{OpenFlamingo} & \textbf{BLIP-2} & \textbf{Kimi-VL} & \textbf{GPT-4o} & \textbf{Gemini 2.0} & \textbf{Average} \\
\midrule
No Defense   & 55.4  & 39.0       & 41.2      & 99.6         & 38.6   & 33.2    & 16.6   & 12.4   & 43.25 \\
Bit-Red~\cite{xu2017feature}      & 51.2  & 37.4       & 38.6      & 99.6         & 36.4   & 28.6    & 17.0   & 13.3   & 41.95 \\
JPEG ~\cite{guo2018countering}        & 34.2  & 26.0       & 28.3      & 74.4         & 27.4   & 23.8    & 8.4    & 8.8   & 29.39 \\
NRP~\cite{naseer2020self}          & 35.8  & 30.4       & 27.0      & 78.6         & 26.6   & 26.8    & 10.0    & 11.2  &  30.80\\
DiffPure~\cite{nie2022DiffPure} & 31.0 & 26.8 & 23.2& 54.0& 25.0&23.6&12.6&10.4 & 25.83\\
\bottomrule
\end{tabular}}
\end{table*}

\subsection{Limitations and Future Works}

Although our proposed method demonstrates strong transferability in the vision-language setting, several limitations remain and open avenues for future research. 
\ding{182} Our current attack focuses on LVLMs and has not yet been extended to broader multimodal systems, such as audio–language models or embodied agents. Generalizing our framework to support additional modalities remains an important future direction.
\ding{183} Our study primarily focuses on untargeted attacks, where the goal is to disrupt the victim LVLM's normal visual understanding and induce incorrect outputs or outputs inconsistent with grounded evidence. 
This threat model captures a broad class of practical risks, including misleading captions, incorrect VQA responses, and perception failures in safety-critical scenarios. 
However, targeted attacks, which aim to elicit a specific attacker-chosen response, represent another important threat model. 
Extending our grounding-driven framework to targeted attacks could be achieved by jointly suppressing the original text-grounded evidence and encouraging alignment with target-specific descriptions or answers. 
We leave a systematic investigation of grounding-driven targeted attacks to future work.
\ding{184} Our evaluation of defenses has been limited to input-level preprocessing methods, which show only marginal effectiveness. These findings underscore the need for exploring stronger defense strategies, such as adversarial training or detection mechanisms, to better safeguard LVLMs against transferable adversarial examples.